\documentclass{mn2e}
\usepackage{graphicx}
\usepackage{times}

\def\arcdeg{\hbox{$^\circ$}}

\def\arcsec{\hbox{$^{\prime\prime}$}}
\def\deg2{\hbox{$\rm deg^{2}$}}

\def\lsim{\mathrel{\rlap{\lower4pt\hbox{\hskip1pt$\sim$}}\raise1pt\hbox{$<$}}}                
\def\gsim{\mathrel{\rlap{\lower4pt\hbox{\hskip1pt$\sim$}}\raise1pt\hbox{$>$}}}                

\begin{document}

\title{Cataclysmic Variables from the Catalina Real-time Transient Survey}

\author[A.J. Drake et al.]{
 A.J.~Drake,$^1$ B.T.~G\"ansicke,$^2$ S.G.~Djorgovski,$^1$ P.~Wils,$^3$ A.A.~Mahabal,$^1$ M.J.~Graham,$^1$
\newauthor T-C.~Yang,$^{1,4}$ R.~Williams,$^1$ M.~Catelan,$^{5,6}$ J.L.~Prieto,$^7$ C.~Donalek,$^1$ S.~Larson,$^8$ and
\newauthor E.~Christensen$^8$\\
$^1$California Institute of Technology, 1200 E. California Blvd, CA 91225, USA\\
$^2$Department of Physics, University of Warwick, Coventry, CV4 7AL\\
$^3$Vereniging voor Sterrenkunde, Belgium\\
$^4$National Central University, JhongLi City, Taiwan\\
$^5$Pontificia Universidad Cat\'olica de Chile, Insituto de Astrof\'isica, 782-0436 Macul, Santiago, Chile\\
$^6$The Milky Way Millennium Nucleus, Santiago, Chile\\
$^7$Department of Astronomy, Princeton University, 4 Ivy Ln, Princeton, NJ 08544, USA\\
$^8$Department of Planetary Sciences, The University of Arizona, 
1629 E. University Blvd, Tucson AZ 85721, USA
}

\volume{000}
\pubyear{0000}

\maketitle
\begin{abstract} 
  
  We present 855 cataclysmic variable candidates detected by the Catalina Real-time Transient Survey (CRTS) of which at
  least 137 have been spectroscopically confirmed and 705 are new discoveries. The sources were identified from the
  analysis of five years of data, and come from an area covering three quarters of the sky. We study the amplitude
  distribution of the dwarf novae CVs discovered by CRTS during outburst, and find that in quiescence they are typically
  two magnitudes fainter compared to the spectroscopic CV sample identified by SDSS.  However, almost all CRTS CVs in
  the SDSS footprint have $ugriz$ photometry. We analyse the spatial distribution of the CVs and find evidence that many
  of the systems lie at scale heights beyond those expected for a Galactic thin disc population.  We compare the
  outburst rates of newly discovered CRTS CVs with the previously known CV population, and find no evidence for a difference
  between them. However, we find that significant evidence for a systematic difference in orbital period distribution.
  We discuss the CVs found below the orbital period minimum and argue that many more are yet to be identified among the
  full CRTS CV sample.  We cross-match the CVs with archival X-ray catalogs and find that most of the systems are dwarf
  novae rather than magnetic CVs.

\end{abstract}

\section{Introduction}

Cataclysmic variables (CVs) are a common state of evolved compact binary systems. Such systems consist of a
main-sequence, sub-giant or brown dwarf star that is filling its Roche lobe and transferring mass onto a white dwarf
(Warner 2003). The accretion process can be either directly onto a strongly magnetic white dwarf or by way of an
intervening accretion disc.  Many CV systems with accretion discs undergo thermal instabilities within their discs
(Meyer \& Meyer-Hofmeister 1981) that give rise to outbursts of up to eight magnitudes, and make up the CV sub-class termed
dwarf novae (e.g. Patterson et al. 1981; Howell et al. 1995). These outbursting events in dwarf novae-type CVs can last 
from days to weeks (e.g. Szkody \& Mattel 1984). Apart from their role in binary star evolution, understanding such
systems is important for cosmology, since CVs remain possible progenitors to type-Ia supernovae explosions (Patat et
al.~2007; Kafka et al.~2012; Immer et al.~2006; Zorotovic et al.~2011).

Historically, the discovery of dwarf nova type CVs has been in large part due to serendipitous detection and subsequent
follow-up studies (G\"ansicke 2005). More recently, confirmation of dwarf nova candidates has been undertaken routinely
by a large network of small telescopes (Kato et al.~2009). The discovery of these systems is aided by the large
intrinsic variations of the sources. However, the lack of deep synoptic wide-field surveys has meant that most
historical CV discoveries have been, either relatively bright nearby CV systems, or fainter systems undergoing very
large outbursts.  An exception to this discovery method has been the Sloan Digital Sky Survey (SDSS), which undertook a
spectroscopic survey of more than a hundred thousand QSO targets (Schneider et al. 2010; Paris et al. 2012).  Due to the
similar optical colours of QSOs and CVs, besides large numbers of QSOs, a few hundred CVs were discovered (Szkody et
al.~2002, 2003, 2004, 2005, 2006, 2007, 2009, 2011). Since these CV systems were identified from quiescent spectra
rather than optical variation, this survey presented an unprecedented insight into the variety of system properties
within the CV population, and led to the firm detection of the predicted accumulation of CVs near the orbital period
minimum (G\"ansicke et al.~2009).  However, as spectroscopic observations require more flux than photometry, the SDSS CV
sample was limited to sources with $i < 19.1$ (although later work followed some targets as faint as $i=20.2$; Richards et al.~2002).

The fact that dwarf novae brighten by many magnitudes during their outbursts enables the discovery of CV systems that
are very faint in quiescence.  However, intrinsically faint systems have lower accretion rates and less frequent
outbursts compared to bright sources, thus introducing a bias in variability-based
searches (Wils et al.~2010).  To find large numbers of the faintest CV systems, it is necessary to repeatedly survey
large areas of the sky.

A number of surveys have begun to systematically explore the astronomical time domain in order to discover optical
transient events, such as CV outbursts.  These projects include the Catalina Real-time Transient Survey (CRTS; Drake et al. 2009a,
Djorgovski et al. 2011), the Panoramic Survey Telescope and Rapid Response System (PanSTARRS; Hodapp et al.~2004), the
Palomar Transient Factory (PTF; Law et al.~2009) and the La Silla Quest survey (LSQ; Rabinowitz et al.~2011).  All these
surveys are capable of discovering hundreds of intrinsically faint CVs during outbursts. However, only CRTS 
openly reports the discovery of CVs.  Future surveys such as SkyMapper (Keller et al. 2007), Gaia (Perryman et al.~2001) 
and the Large Synoptic Survey Telescope (LSST; Ivezic et al.~2008) are also expected to detect numerous CVs.

In this paper we describe the CV systems that were detected by CRTS in data taken by the Catalina Sky Survey 
between 2007 November 8th and 2012 July 31st. We will then investigate the basic properties of these systems and
outline areas where additional work is required to better understand their nature.

\section{Observational Data}

The Catalina Sky Survey\footnote{http://www.lpl.arizona.edu/css/} began in 2004 and uses three telescopes to repeatedly
survey the sky between declination $\delta = -75$ and +65 degrees in search of Near-Earth Objects (NEOs) and Potential
Hazardous Asteroids (PHAs).  In addition to asteroids, all the Catalina data is analyzed for transient sources by 
CRTS (Drake et al.~2009a; Djorgovski et al 2011).

Each of the survey telescopes is run as separate sub-surveys.  These consist of the Catalina Schmidt Survey (CSS), the
Mount Lemmon Survey (MLS) and the Siding Spring Survey (SSS).  In this paper we analyse data taken by all three
telescopes, namely the 0.7m CSS telescope and the 1.5m MLS telescope in Tucson, AZ and the 0.5m (SSS) Uppsala Schmidt
at Siding Spring Observatory, Australia. Transient processing of CSS data by CRTS began on 2007 November 8th, 
while for MLS data it began on 2009 November 6th, and for SSS on 2010 May 5th.

Each telescope currently has a $4{\rm k} \times 4{\rm k}$ CCD camera, which for
the CSS, MLS and SSS cover 8.2, 1.1 and 4 square degrees, respectively.  In general each telescope avoids the Galactic
latitudes less than 10 to 15 degrees due to reduced source recovery in crowded stellar regions. Because of its smaller
field-of-view the MLS 1.5m telescope predominately observes only ecliptic latitudes $-10\arcdeg < \beta < 10 \arcdeg$,
whereas the CSS covers $-30\arcdeg < \delta < 65 \arcdeg$ and the SSS mainly covers $-75\arcdeg < \delta < 0 \arcdeg$.
In total, $\sim$ 30,000 $\rm deg^2$ of the sky are surveyed by the three telescopes.

Observations are taken during the darkest 21 nights per lunation in sets of four images each separated by 10 minutes.
Exposures typically last for 30 seconds. All images are taken unfiltered to maximize throughput. Photometry
is carried out using the aperture photometry program SExtractor (Bertin \& Arnouts 1996) and is transformed
to $V$ using standard stars as noted by Drake et al.~(2013).

\section{Transient Selection}

The CRTS project is aimed at the discovery of astrophysical sources such as CVs and supernovae that undergo transient
brightness variations. Details of the transient detection procedures in CRTS are given by Drake et al.~(2009a).  In
short, transient sources are identified by comparing them with detections in past photometric catalogues including
CSS, USNO (Monet et al. 2003) and SDSS DR8 (Aihara et al. 2011). For selection as a transient, an object must be a 
point source. It also must either not be present in previous catalogs, or be present and have varied with a high significance.

Objects detected within archival catalogues qualify as transients in CRTS when they increase in brightness by at least
three times the observed scatter at the source brightness provided there is a minimum increase of 0.65 magnitudes.
Transient sources must be present in at least three of the four observations within a set of observations. The 0.65 mag
threshold is a significant change from Drake et al.~(2009a), and was adopted after careful consideration of transient
discoveries and the variations of common periodic variable stars. For example, almost all of the RR Lyrae detected by
Drake et al.~(2013) in Catalina data vary by less than 1.3 magnitudes peak-to-peak. Therefore, they generally fall below
the 0.65 magnitude threshold when compared to their median magnitudes.  Among the $\sim6200$ transient sources
discovered by CRTS before July 2012, only a couple of dozen periodic variable stars (mainly halo LPVs) have met the
transient detection criterion, thereby demonstrating that this threshold is effective in both removing periodically
variable stars as well as catching low-amplitude transients.

In addition to the variability threshold, transient selection requires a number of filters to remove artifacts and
artificial variations.  Additional filters are run on the images containing transient candidates to remove objects that
are detected as transients, but were missing from the input catalogs due to source blending and bad image data.  Moving
objects such as asteroids and comets are analysed by the CSS NEO survey and are removed based on known Minor Planet
Center sources as well as based on any motion over the 30 minutes between sets of four observations.  Transient
candidates are also compared with sources from archival data from the SDSS, the Digital Sky Survey (DSS), and the
Palomar-Quest (PQ) survey in order to classify the sources and remove any possible remaining artifacts. All the transient
candidates that pass the filtering stages are visually inspected and classified using their lightcurves along with
information from archival surveys.

In the case of CV transients, there are a number of factors that can affect the detection sensitivity in various ways. For
example, the efficiency of detecting frequently outbursting CVs is lower than one might expect. This is because objects
that were in outburst in the comparison catalogs cannot reach the detection threshold. Many of the CVs detected
by Wils et al.~(2010) were in outburst in USNO data and thus would not meet our transient criteria. Furthermore, bright
CVs may become saturated during outburst and go undetected, while in contrast, CVs with faint quiescent magnitudes
can only be detected during very large outbursts.


\section{The CRTS Cataclysmic Variables}

\subsection{CV Candidates}

Among the $\sim$ 6,200 CRTS transient sources detected before 2012 July 31st, 1062 were identified as CV candidates.
The classification as CV candidate relies on several different lines of evidence. Transients that exhibited
outbursts in prior CSS, PQ, DSS or other archival sources were deemed to be very good CV candidates and so were
objects with blue PSF-like counterparts in SDSS or GALEX photometry (provided there was no radio flux present). 

The presence of photometric variations in follow-up data and Balmer emission lines in spectroscopic observations was
also used to classify some of the systems as good CV candidates.  In contrast, if no prior outburst was seen, or the
colour and extent of the quiescent source was unclear, the transient source was deemed ambiguous. These objects are
rejected from our sample of CV candidates.  This selection excludes 235 sources with significant outbursts, but
no additional evidence that they were CVs.
The remaining 855 objects are hereafter identified as good CV candidates.  Among these sources, 150
systems were known to be CVs prior to their detection by CRTS.  Ongoing photometric and spectroscopic follow-up of CRTS
CV candidates suggests that our selection process has $>95\%$ accuracy at identifying CV systems (eg. Kato et al.~2009,
2010, 2012a, 2013; Woudt, et al. 2012; Thorstensen \& Skinner ~2012).


\begin{table*}
\label{CRTSspectra}
\caption{CRTS Cataclysmic Variables}
\begin{minipage}{186mm}
\begin{center}
\begin{tabular}{@{}lcllllll}
\hline 
CRTS ID & RA & Dec (J2000) & $\rm Mag_{O}$ & $\rm Mag_{Q}$ & $\rm N_{det}$ & Prior ID & Reference\\
\hline 
CRTS\_J000024.7+332543 & 00:00:24.67 & +33:25:43.0 & 15.7 &  & 2 & CSS100910$:$000025+332543 & \\
CRTS\_J000130.5+050624 & 00:01:30.47 & +05:06:23.6 & 15.3 & 20.2 & 2 & CSS101127:000130+050624& \\
CRTS\_J000659.6+192818 & 00:06:59.59 & +19:28:17.8 & 17.5 & $>$21.0 & 2 & CSS100602:000700+192818 & \\
CRTS\_J000720.7+200722 & 00:07:20.74 & +20:07:21.7 & 17.1 &  & 1 & CSS110921:000721+200722 & \\
CRTS\_J000938.2-121017 & 00:09:38.25 & -12:10:16.7 & 14.5 &  & 1 & CSS101007:000938-121017 & \\
CRTS\_J000945.5+402928 & 00:09:45.46 & +40:29:28.0 & 17.9 &  & 1 & CSS091120:000945+402928 & \\
CRTS\_J001019.3+410455 & 00:10:19.31 & +41:04:54.9 & 15.6 & 19.5 & 1 & CSS091120:001019+410455 & \\
CRTS\_J001158.3+315544 & 00:11:58.28 & +31:55:44.0 & 17.3 &  & 1 & CSS101111:001158+315544 & \\
CRTS\_J001310.6+212108 & 00:13:10.63 & +21:21:8.1 & 18.0 & $>$20.4 & 3 & CSS101201:001311+212108 & \\
CRTS\_J001339.6+332124 & 00:13:39.58 & +33:21:23.5 & 17.9 & $>$19.9 & 2 & CSS101111:001340+332124 & \\
CRTS\_J001449.5-523215 & 00:14:49.55 & -52:32:15.4 & 16.9 &  & 1 & SSS110916:001450-523215 & \\
CRTS\_J001538.3+263657 & 00:15:38.26 & +26:36:56.8 & 13.3 & 17.7 & 2 & CSS090918:001538+263657 & \\
CRTS\_J001636.9+185615 & 00:16:36.88 & +18:56:15.2 & 18.6 &  & 2 & CSS080202:001637+185615 & \\
CRTS\_J001828.4+215519 & 00:18:28.36 & +21:55:19.4 & 18.7 & $>$21.5 & 1 & CSS101107:001828+215519 & \\
CRTS\_J001952.2+433901 & 00:19:52.24 & +43:39:1.4 & 15.6 & & 1 & CSS120131:001952+433901 & \\
\hline
\end{tabular}
\end{center}
\end{minipage}
\medskip 
The full table will be available in the online version.
Col. (1) presents the CRTS identifier.
Cols. (2) \& (3). present the right ascension and declination, respectively.
Col. (4) presents the observed peak V magnitude.
Col. (5) presents the CRTS quiescent magnitude or limit (if known).
Col. (6) presents the number of outbursts detected by CRTS.
Col. (7) presents the detection ID from CRTS or other prior IDs when previously known.
Col. (8) presents the reference for spectroscopic follow-up from the following list.
$\rm^{I}$Szkody et al.~(2002-2011), 
$\rm^{II}$Thorstensen \& Skinner (2012), 
$\rm^{III}$This work, 
$\rm^{IV}$Quimby et al.~(2008),
$\rm^{V}$Levitan et al.~(2013),
$\rm^{VI}$Garnavich et al.~(2012),
$\rm^{VII}$Breedt et al.~(2012),
$\rm^{VIII}$Woudt et al.~(2012),
$\rm^{IX}$Wright et al.~(2012),
$\rm^{X}$Wils et al.~(2010),
$\rm^{XI}$Croom et al.~(2001),
$\rm^{XII}$Jones et al.~(2004).
$\rm^{XIII}$Drake et al.~(2009a).

\end{table*}

In Table 1, we present the parameters of the 855 good CV candidates detected by CRTS before 2012 August 1st.  The table
presents the number of times the source was detected in outburst by CRTS up to this date.  A large fraction of the
objects have additional outbursts in Catalina photometry that was taken before CRTS began searching these data for
transients.  For completeness, we have included the 64 CV candidates discovered between 2007 November 8th and 2008 May
14th that were presented by Drake et al.~(2009a). Many of these systems have now had additional outbursts and a few have
been spectroscopically confirmed.  We also include the 150 previously known sources, many of which are well
characterized by by published follow-up observations.  The detection of these sources demonstrates the overall
sensitivity of CRTS.


\begin{figure}{
\includegraphics[width=84mm]{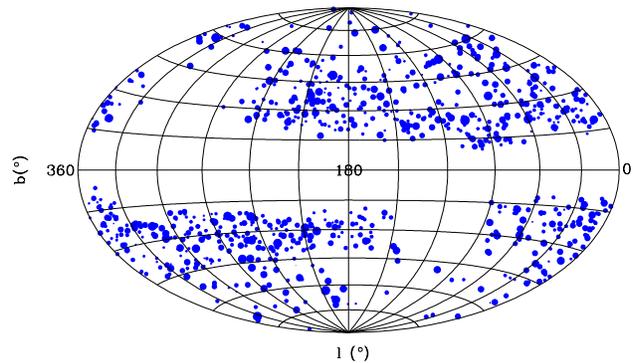}
\caption{\label{Ait}
The distribution of CRTS CV candidates in Galactic
coordinates (Aitoff projection). The radii of the points are proportional to the
peak magnitude with the brightest points largest. Gaps are present in
the regions not observed by Catalina, i.e., near the celestial poles and in the galactic plane. 
}
}
\end{figure}

\begin{figure}{
\includegraphics[width=84mm]{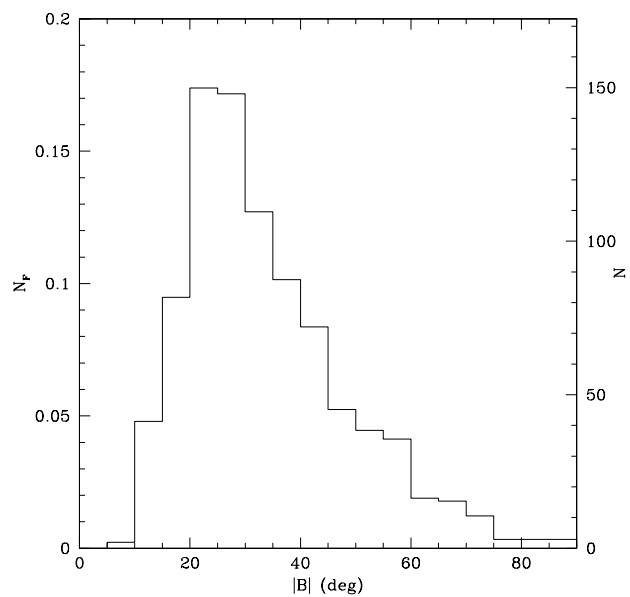}
\caption{\label{Lat}
The Galactic latitude distribution of CRTS CV candidates. 
}
}
\end{figure}

In Figure \ref{Ait}, we present the Galactic distribution of CV candidates and outburst brightnesses.  In Figure
\ref{Lat}, we plot the distribution of these sources as a function of Galactic latitude. As expected, the number of
sources drops significantly at high Galactic latitudes, whereas near the Galactic plane few sources are detected 
due to reduced survey coverage.

\subsection{Other outbursting transients}

In addition to CV systems, the outbursting optical transients discovered by CRTS also include blazars and
supernovae.  Most known blazars have been discovered as flat-spectrum sources in radio survey data (Healey et al.~2007;
Healey et al.~2008).  Unlike blazars, CVs are generally not radio sources with a few notable exceptions (eg. Mason \& Gray 2007;
Kording et al. 2008; 2011). Thus CVs and blazars can be separated based on radio data.  For this purpose we have used observations
from the NRAO VLA Sky Survey (NVSS; Condon et al.~1998), the Faint Images of the Radio Sky at Twenty-cm (FIRST; Becker
et al.~1995), and the Sydney University Molonglo Sky Survey (SUMSS; Mauch et al.~2003). The entire survey region of CRTS
is covered by one or more of these radio surveys. The NVSS survey covers the sky at 1.4 GHz in the region $\delta >
-40\arcdeg$.  The SUMSS survey covers the entire sky at latitudes $\delta < -30\arcdeg$ for $|b| > 10\arcdeg$ at 843 MHz, and the
FIRST survey covers the 10,000 square degrees around the Galactic poles at twenty centimetres. To reduce the possibility
of misclassifying blazar sources as CVs, we routinely inspected the images generated by these surveys.

\begin{figure}{
\includegraphics[width=84mm]{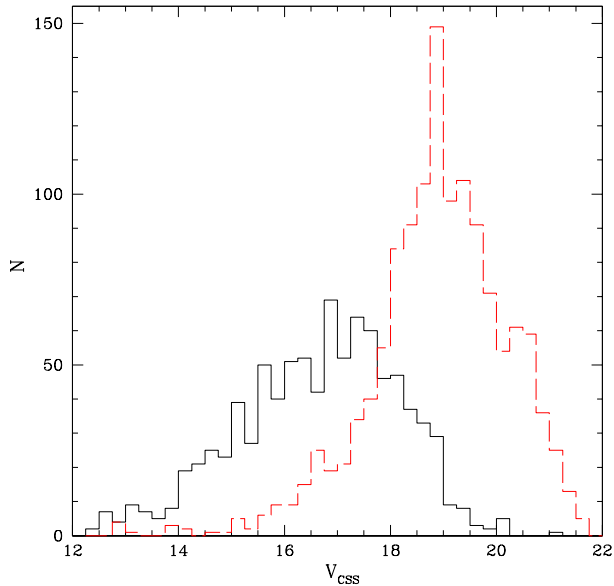}
\caption{\label{Hist}
A histogram of the detection magnitudes of CRTS CVs and SN.
CVs are given by the solid-line, while SN are given by the
dashed-line.
}
}
\end{figure}

In contrast to CVs, supernova explosions give rise to a single brightening event that typically lasts for a few months.
The host galaxies of supernova explosions are usually visibly extended in archival images. Furthermore, supernovae, like
blazar outbursts, are roughly isotropically distributed over the sky. Both supernovae and blazars have different
quiescent colours than CVs (i.e. which tend to be blue). Lastly, supernovae generally give rise to smaller increases in
brightness relative to the brightness of their host galaxies and have fainter apparent magnitudes.  To demonstrate this,
we present a histogram of the CV and supernova detection magnitudes in Figure \ref{Hist}.  The distribution of CV
outburst magnitudes is offset towards brighter apparent magnitudes from supernova explosions.  The incidence of bright 
supernova explosions is limited to a small number of events that occur each year in nearby galaxies.

\section{CV Activity}

\begin{figure}{
\includegraphics[width=84mm]{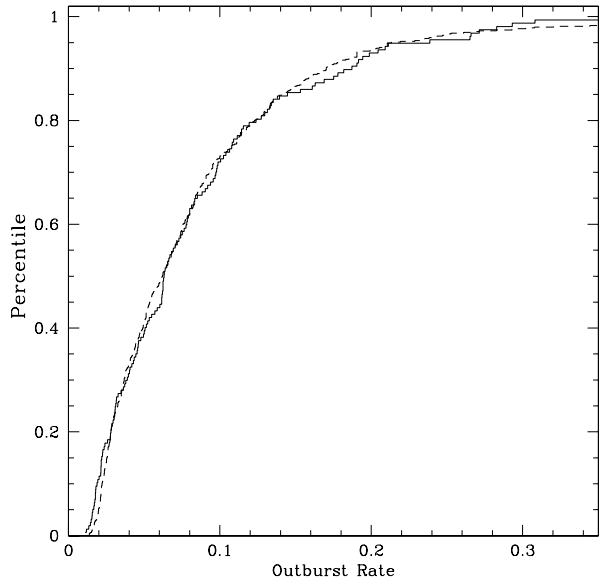}
\caption{\label{Rate}  
  The cumulative distribution of CV outburst rates. We define the outburst rate as the ratio of the number 
  of nights observed in outburst to the number of nights of observation. The previously known CVs are given 
  by the solid-line, whereas the new sources are given by the dashed-line.
  } }
\end{figure}

\begin{figure*}{
\includegraphics[width=150mm]{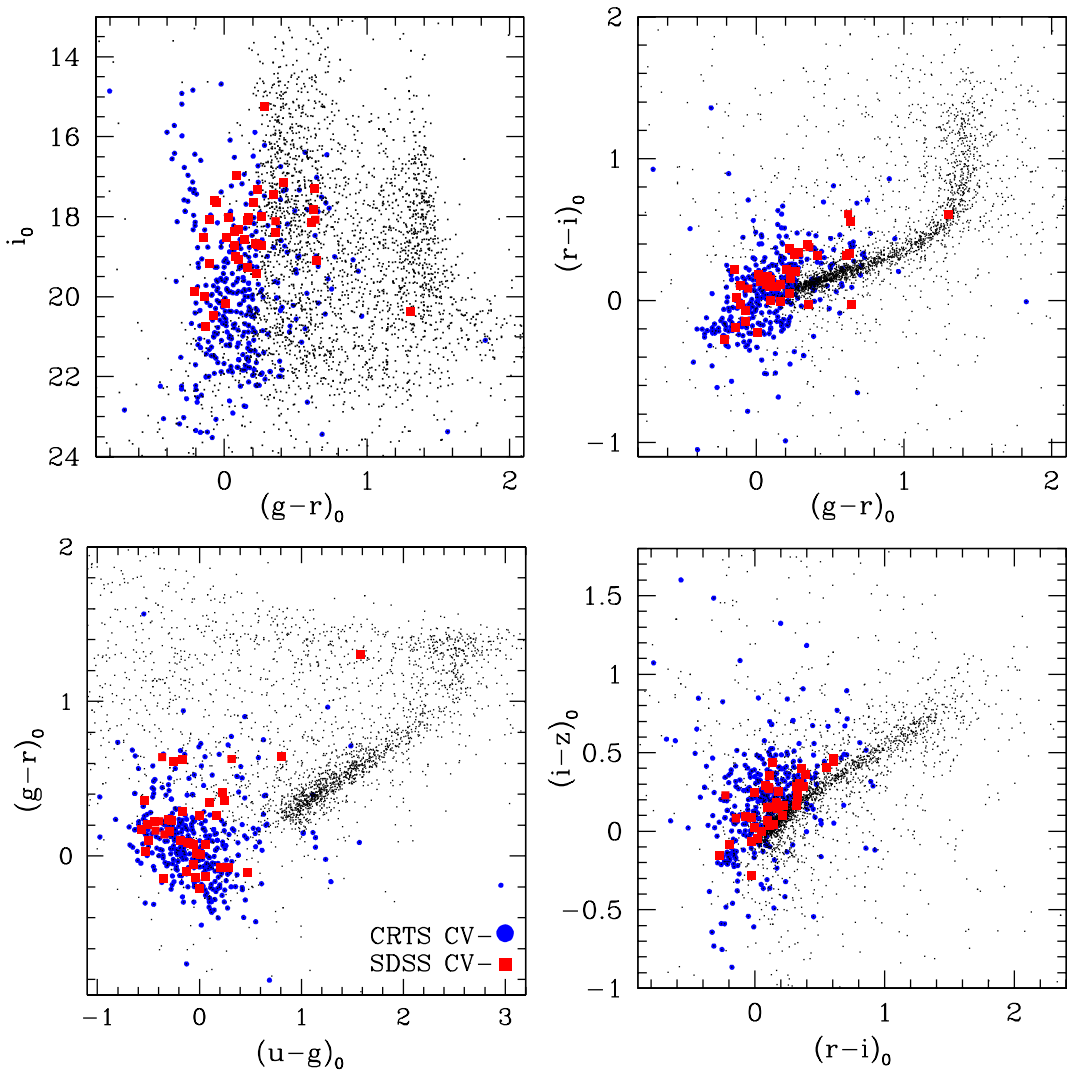}
\caption{\label{SDSS}
The colour and brightness distribution of CRTS CV candidates.
CV candidates are large blue dots. Red squares are sources with SDSS
spectra. The black points are the point sources (stars, QSOs and other unresolved sources) 
from SDSS that lie within an arc minute of each CV candidate.
}
}
\end{figure*}

In order to investigate the CV outburst frequency of the candidates, we divided the number of outbursts detected by the
number of nights the sources were observed. As most locations are only sampled every few weeks, individual outbursts are
generally only observed on one night. Therefore, this ratio gives the approximate outburst rate. In Figure \ref{Rate},
we plot the cumulative distibution of rates for the entire set of CVs detected by CRTS. We separate the previously known
CVs from new discoveries to investigate whether the CV outburst rates vary because of differences in selection, etc. 
Performing a KS-test on the two sets of rates we find a $D$-statistic of 0.068, corresponding to a $P$-value of 0.58.
Therefore, we find no significant evidence for a difference in the distributions of outburst rates for the known 
(and generally brighter) CVs compare to the newly discovered CVs.  This result is in slight disagreement with the
findings of G\"ansicke et al.~(2009), based on SDSS CVs, and the results of Wils et al.~(2010). However, we do find 
a very large number of CVs in the sample have low outburst rates in agreement with these authors. This is reflected 
in the fact that 55\% of the good CV candidates have only been detected in outburst once.  There is insufficient 
data to quantify rates for the sources where outbursts occur $< 3\%$ of the time.


\section{Comparison with SDSS data}

\begin{figure}{
\includegraphics[width=84mm]{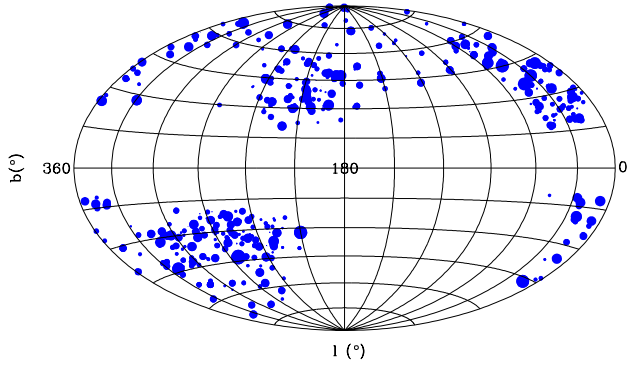}
\caption{\label{DistSDSS}
The Galactic coordinates of CRTS CV candidates with SDSS photometry (Aitoff projection). 
The radii of the points are proportional to their SDSS $i$-band magnitudes. The brightest 
sources are largest. 
}
}
\end{figure}

As CSS images are taken without filters there is very little information regarding the colours of candidates. However, the
SDSS DR8 dataset provides photometry in five filters and spans approximately half of the area covered by CRTS (14,555
$\rm deg^2$).  We cross-matched the CV candidates with the SDSS DR8 photometric database and found 416 common objects.
In Figure \ref{SDSS}, we present the colours and magnitudes of the CV candidates with SDSS photometry. 
As expected, the CV candidates detected by CRTS have significantly bluer $(u-g)_0$ colours than normal main-sequence stars.
The distribution of SDSS $i_0$ magnitudes clearly shows that the CVs discovered by CRTS are much fainter than the SDSS
spectroscopic CV sample. Since CVs are in outburst for only a small fraction of the time, in most cases the SDSS
photometry was taken when the CVs are in quiescence. However, by comparing CSS magnitudes with SDSS ones, there are a
few cases where SDSS imaging was clearly taken during an outburst. In Figure \ref{DistSDSS}, we present the spatial
distribution of CRTS CVs in SDSS $i$-band photometry and apparent brightness.

\begin{figure}{
\includegraphics[width=84mm]{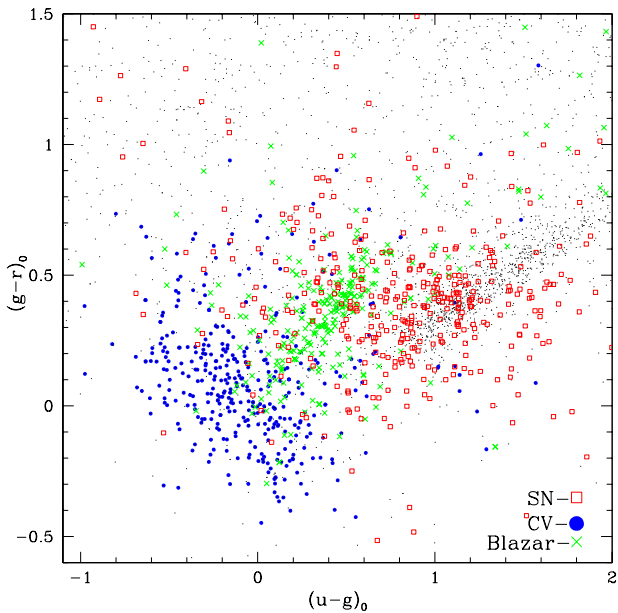}
\caption{\label{SNBlaz}
The colours of CRTS CV candidates compared with other transient sources. 
CVs are large blue dots. Red boxes are supernova hosts.
Green crosses are Fermi and CRTS blazars. Small black dots are point sources 
within one arc minute of each CV.
}
}
\end{figure}

In Figure \ref{SNBlaz}, we compare the SDSS colours of CVs with those of supernova hosts galaxies and blazars.  For
the blazars, in addition to the presence of radio emission, we can see that these sources are generally redder than 
CVs, both in their $(g-r)_0$ and the $(r-i)_0$ colours. They also form quite a tight group. The supernova hosts have a 
much more diverse range of colours than CVs and blazars, and are also clearly redder than quiescent CVs.

As SDSS photometry is much deeper than Catalina data, SDSS photometry enables us to assess quiescent source colours and
magnitudes to levels below our detection threshold.  In addition, a comparison between Catalina $V$ magnitudes 
and SDSS photometry reveals a relatively close match to SDSS $i$-band magnitudes. This is likely due to the red sensitivity of
unfiltered Catalina data. We investigated the transformation of the SDSS
photometry\footnote{http://www.sdss.org/dr7/algorithms/sdssUBVRITransform.html} and found poorer results.

\begin{figure}{
\includegraphics[width=84mm]{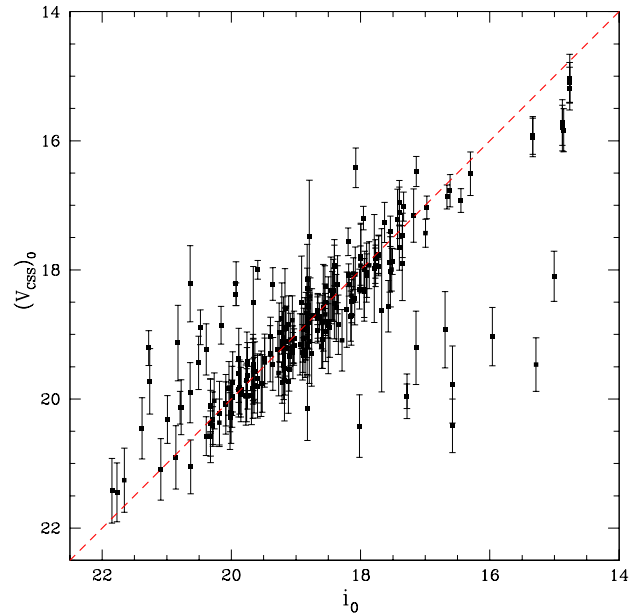}
\caption{\label{Cal}
A comparison between the CSS $V$ and SDSS $i$ magnitudes, for the 220 CV candidates 
with SDSS photometry and a measurement of quiescent magnitude in Catalina data.
}
}
\end{figure}

In Figure \ref{Cal}, we compare the Catalina $V$ magnitudes with SDSS $i$-band photometry.
We only compare the 220 candidates that have both a Catalina quiescent magnitude values and
SDSS photometry. A number of the SDSS magnitudes are significantly offset from the Catalina measurements.  As noted
above, some of the SDSS sources were imaged while they were in outburst. For a few sources the SDSS magnitudes
are much fainter. This is likely to be due to the Catalina images having lower resolution than SDSS.
Multiple SDSS sources can contribute to a single CSS object and thus make CSS source appear brighter.


For comparison purposes we note that the photometric uncertainties of individual CSS observations vary from $0.03 - 0.5$
magnitudes. However, the measured dispersion in measurements for the CV is typically much larger than this because of
variability.  For the SDSS photometry $\sigma_i < 0.1$ magnitudes for most sources.  After removing the outliers that
typically differ by more than two sigma (1.1 mag), we were left with a sample of 184 CVs with SDSS and Catalina
magnitudes.  For these objects we find an average difference between $i$-band and Catalina $V$ of -0.01 magnitudes with
$\sigma=0.33$ magnitudes. This level of scatter is much larger than derived by Drake et al.~(2013) for
horizontal-branch stars observed by both Catalina and SDSS. We believe this is due to two main sources, namely
variations between the SEDs of the CVs, and the intrinsic variability of CVs.

\begin{figure}{
\includegraphics[width=84mm]{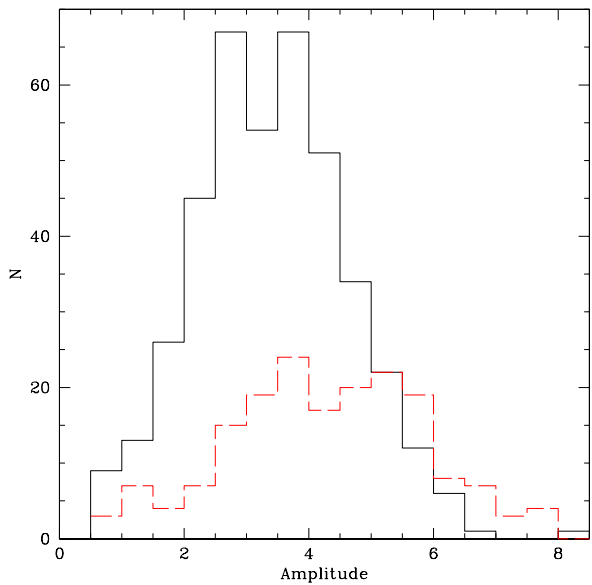}
\caption{\label{HistAmp}
Histograms of the outburst amplitudes for CRTS CVs.
The solid line presents the distribution for 416 sources 
where the quiescent magnitude was bright enough to measure 
in Catalina data. The dashed-line shows the distribution
for 196 fainter sources where their SDSS $i$-band 
magnitude is assumed to be equivalent to the $V_{\rm CSS}$ 
quiescent brightness.
}
}
\end{figure}

By comparing the SDSS $i$-band magnitudes with the Catalina outburst magnitudes we can estimate the outburst amplitudes
of faint sources.  In Figure \ref{HistAmp}, we present histograms of the CV outburst amplitudes.  For bright CVs we
measure the difference between the average quiescent brightness ($M_Q$) and the average peak measured brightness
($M_O$). For CVs with faint quiescent states, and SDSS photometry, we determine the difference between the peak outburst
magnitude and the SDSS $i$-band magnitude. CVs with fainter apparent magnitudes clearly exhibit, on average larger
amplitude outbursts, i.e. these faint CVs exhibit a large number of outbursts greater than 5.5 magnitudes.  However, this
distribution is subject to selection effects, as only large amplitude outbursts can be detected by CRTS for faint CVs.
Furthermore, some of the CVs detected in quiescence in CSS data, may be missed when they become saturated during large
amplitude outbursts.

\begin{figure}{
\includegraphics[width=84mm]{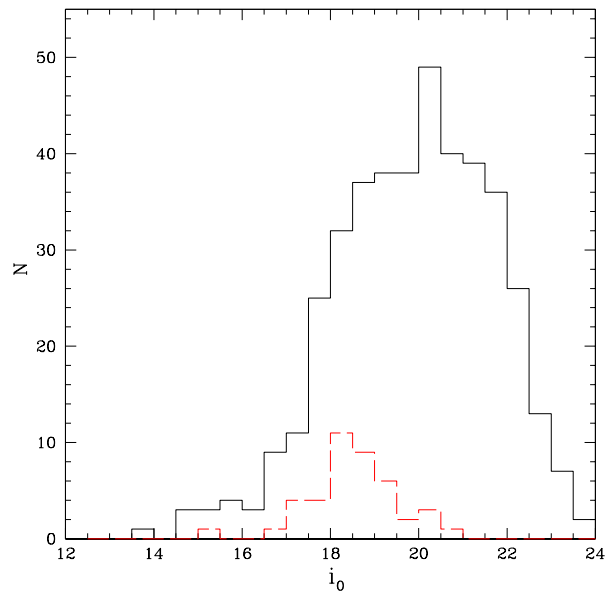}
\caption{\label{HistHost}
The distribution of SDSS $i$-band magnitudes for CRTS CVs.
All CVs with SDSS photometry are given by the solid-line, 
while CVs that have spectra taken by SDSS are given by 
the dashed-line.
}
}
\end{figure}

In Figure \ref{HistHost}, we present the distribution of SDSS $i$-band magnitudes.  The figure demonstrates how the CRTS
CV detections go well beyond the limits of SDSS spectroscopy and thus probe a larger volume. In addition, it shows that
many of the bright CVs found by CRTS within the SDSS imaging fields do not have SDSS spectra. These sources are therefore
missing from the SDSS CV catalogs (Szkody et al. 2002-2011).  However, SDSS spectroscopy involved complex target
selections largely aimed at QSOs (Richards et al.~2002) and BHB stars (Yanny et al. 2009), rather than CVs.  It is also
possible that a number of these sources come from areas where spectroscopy was either not undertaken or completed by
SDSS.

SDSS obtained spectra of 285 CVs (Szkody et al.~2002-2011) and of these 184 were new discoveries. Among the 285 CVs from
SDSS, 33 are classified as polars, 7 as intermediate polars, 6 as Nova-like, and 70 dwarf novae. The
remaining 169 do not have clear type classifications.  CRTS covers almost all of the area observed by SDSS and is mainly
sensitive to dwarf novae. Therefore, assuming the overall fraction of dwarf novae in SDSS is the same as among the classified
fraction (60\%), we would expect to have detected 170 of the SDSS CVs as dwarf novae having outbursts in CRTS (assuming 
100\% detection completeness).
However, only 64 SDSS CVs were detected by CRTS during outburst. This suggests that only a third of the dwarf novae
in the SDSS area have been classified as such. The total fraction of dwarf novae in the SDSS spectroscopic sample is
likely to be higher since many of the systems have not been monitored extensively, but the above numbers suggest that 
many of the SDSS CVs suspected to be dwarf novae have rare outbursts. Clearly a large number of dwarf novae are yet 
to be detected within the SDSS coverage area.

\section{Photometric follow-up}

Since the CRTS survey began, all optical transient discoveries have been made public within minutes of 
their detection. Rapid follow-up is particularly important for CVs since the outbursts may only last for 
a few days. As CSS only observes the same sources every two to four weeks, and CVs typically have shorter 
rise times than decline times (Cannizzo et al. 2010), outbursts were caught more often declining than rising when they were
discovered.  Notification of transient detections was made by way of project web pages.  Additionally, since 2009
September 21st, every new CRTS transient discovery included a {\it CRTS Circular} that was linked to webpages as well as posted
to the SkyAlert service (Williams et al.~2010) and the LSST/CRTS iPhone alert app\footnote{http://www.lsstcorp.org/transientevents/index.html}. 
In some cases, notifications were also sent as Astronomer's Telegrams (ATels, eg. Drake et al. 2009b, 2009c, 2009d, 2011a).

Following the first spectroscopically confirmed CV discovered by CRTS (Djorgovski et al.~2008), photometric follow-up
began with the Variable Star Network project (VSNet; Kato et al.~2009; 2010; 2012a; 2013).  Projects undertaken by VSNet largely involve
photometric time-series of bright CVs in outburst with small telescopes reaching limiting magnitudes of $V=16$ to 17. 
To enable such follow-up we created specific webpages (one for each Catalina telescope) where CVs brighter than $V=17$ 
were posted as soon as they were classified\footnote{http://nesssi.cacr.caltech.edu/catalina/BrightCV.html}.
Additionally, since some CVs exhibit lower-amplitude variability that is 
missed by our detection software, we also created a watch-list based on the Ritter \& Kolb~(2003) catalog of known CVs.
For each of the objects observed by CSS data we extracted the existing archival photometry. Then we set up software to 
automatically update the light curves of those known CVs whenever they were
observed\footnote{http://nesssi.cacr.caltech.edu/catalina/CVservice/CVtable.html}.  Since known CVs require no detection
filtering these updates occur immediately.  This service produces a snapshot of the activity level for more than 1,000 CVs
with a couple of hundred known CVs covered each night. Notably, Mukadam et al.~(2011) used this information to discovered the
only known outburst of CV SDSS J074531.92+453829.6.  Similarly, Southworth et al.~(2009) used the data to find and
constrain the rate of outbursts for eclipsing CV SDSSJ100658.40+233724.4.

Between 2008 August 4th and 2011 October 25th the VSNet project sent follow-up requests for 132 bright CVs
discovered by CRTS.  Of these, 57 were followed sufficiently to determine that they are SU UMa-type dwarf novae, and
their superhump periods were used to estimate orbital periods (Kato et al.~2009; 2010; 2012a; 2013).  Additionally, Woudt \& Warner (2010) and Woudt et al.
(2012) together report high-speed photometric observations of 22 CRTS CVs. Woudt et al.~(2012) note that 115 CRTS CVs
have had periods determined and almost all of these lie below the CV period gap.  Wils et al.~(2011) combined SDSS,
GALEX and CRTS data along with astrometric catalogs to discover 64 new CVs. Their results indicated that besides systems
that are faint because they are farther away, there also exists a population of intrinsically faint dwarf novae with
rare outbursts. This result was supported by Thorstensen \& Skinner (2012).

\section{CV Spectra}

Of the 855 {\em good} CV candidates detected by CRTS, more than 137 have already been spectroscopically confirmed.  This
number excludes historical CVs which have largely already been followed spectroscopically (eg. TY PsA; Warner, et al.
1989, and TT Boo; Bruch \& Schimpke 1992).  As noted above, 64 of these systems were either discovered, or
spectroscopically confirmed by the SDSS (Szkody et al. 2002-2011). Recently, 36 additional systems were spectroscopically
confirmed by Thorstensen \& Skinner~(2012).  Additionally, CRTS CVs have been confirmed by Woudt et al.~(2012), Wright
et al.~(2012), Levitan et al.~(2013), Breedt et al.~(2012), and Littlefield et al.~(2013).  

From our own observations we have confirmed 33 CRTS CV candidates using spectroscopy with the Palomar 5m (P200), Keck, 
and SMARTS telescopes. All of these sources were followed during regular spectroscopic confirmation of CRTS optical 
transients. Since our follow-up targets were generally transient sources of uncertain nature, the CVs we observed are 
mainly sources where there was no evidence for a blue point source within archival images from SDSS or DSS.
In Table 3, we present the CVs that were confirmed during our CRTS follow-up. 
The table is presented in chronological order of discovery.

\begin{table}
\label{spectra}
\caption{CVs confirmed by CRTS during transient follow-up.}
\begin{tabular}{@{}lcl}
\hline 
Detection ID & Telescope+Instrument & Reference\\
\hline 
CSS080130:021110+171624 & P200+DBSP    &   I,II,III,IV\\
CSS080227:112633-100210 & P200+DBSP    &   I\\
CSS080505:163121+103134 & P200+DBSP    &   V\\
CSS080606:164147+121026 & P200+DBSP$\ast$   &  VI\\
CSS080606:162322+121334 & P200+DBSP    &   \\   
CSS081026:023839+355648 & P200+DBSP$\ast$   & \\
CSS090416:164413+054158 & P200+DBSP$\ast$   & \\
CSS090826:223958+231837 & Keck+LRIS     &  VII \\
CSS090910:223418-035530 & SMARTS+RCSpec &  VIII \\
CSS090917:221344+173252 & SMARTS+RCSpec &  VIII\\
CSS091024:042229+161430 & Keck+LRIS$\dagger$ & \\
CSS100108:081031+002429 & P200+DBSP & \\  
CSS100313:085607+123837 & P200+DBSP & \\ 
CSS100911:022648+152539 & P200+DBSP & \\
CSS100916:215226-012850 & Keck+LRIS & \\ 
CSS110114:091937-055519 & P200+DBSP & \\ 
CSS110208:135717-093238 & P200+DBSP$\ast$ & \\
MLS100313:131245-064047 & P200+DBSP & \\
CSS100507:164354-131525 & Keck \& P200 &  VII\\ 
MLS101203:050253+171041 & P200+DBSP & \\
CSS110501:094825+204333 & P200+DBSP & \\  
CSS110610:134306+520843 & P200+DBSP$\dagger$ & \\
CSS110623:173517+154708 & P200+DBSP & \\  
CSS110628:220857+200440 & P200+DBSP$\ast$ & \\
CSS111003:054558+022106 & Keck+LRIS & \\
CSS111022:205145+075305 & Keck+LRIS & \\ 
CSS111027:075959+211936 & P200+DBSP & \\
CSS120526:165741-055625 & Keck+LRIS & \\
CSS120113:040822+141516 & P200+DBSP & \\ 
CSS120610:135419+273603 & Keck+LRIS & \\ 
CSS120612:173245+094746 & Keck+LRIS & \\ 
CSS120613:212655-012054 & Keck+LRIS & \\
\hline
\end{tabular}
\medskip 
Objects marked with '$\ast$' have featureless spectra.
Objects marked with '$\dagger$' have noisy identifying spectra.
$\rm^{I}$Glikman et al.~(2008), 
$\rm^{II}$Djorgovski et al.~(2008), 
$\rm^{III}$Drake et al.~(2009a), 
$\rm^{IV}$Kato et al.~(2009),
$\rm^{V}$Mahabal et al.~(2008),
$\rm^{VI}$Djorgovski et al.~(2009)
$\rm^{VII}$Djorgovski et al.~(2010).
$\rm^{VIII}$Drake et al.~(2009b).
\end{table}

\begin{figure*}{
\includegraphics[width=84mm]{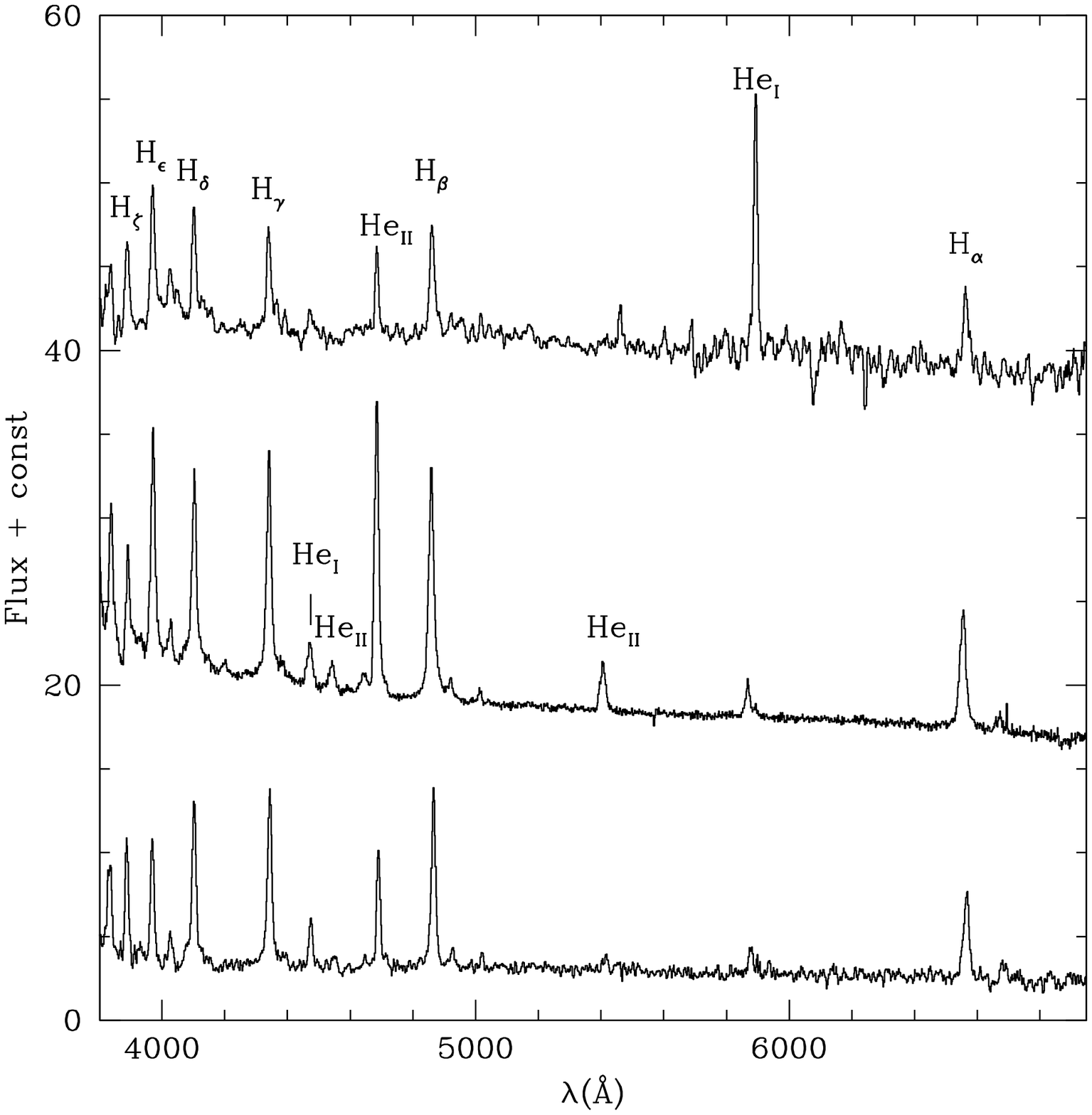}\includegraphics[width=84mm]{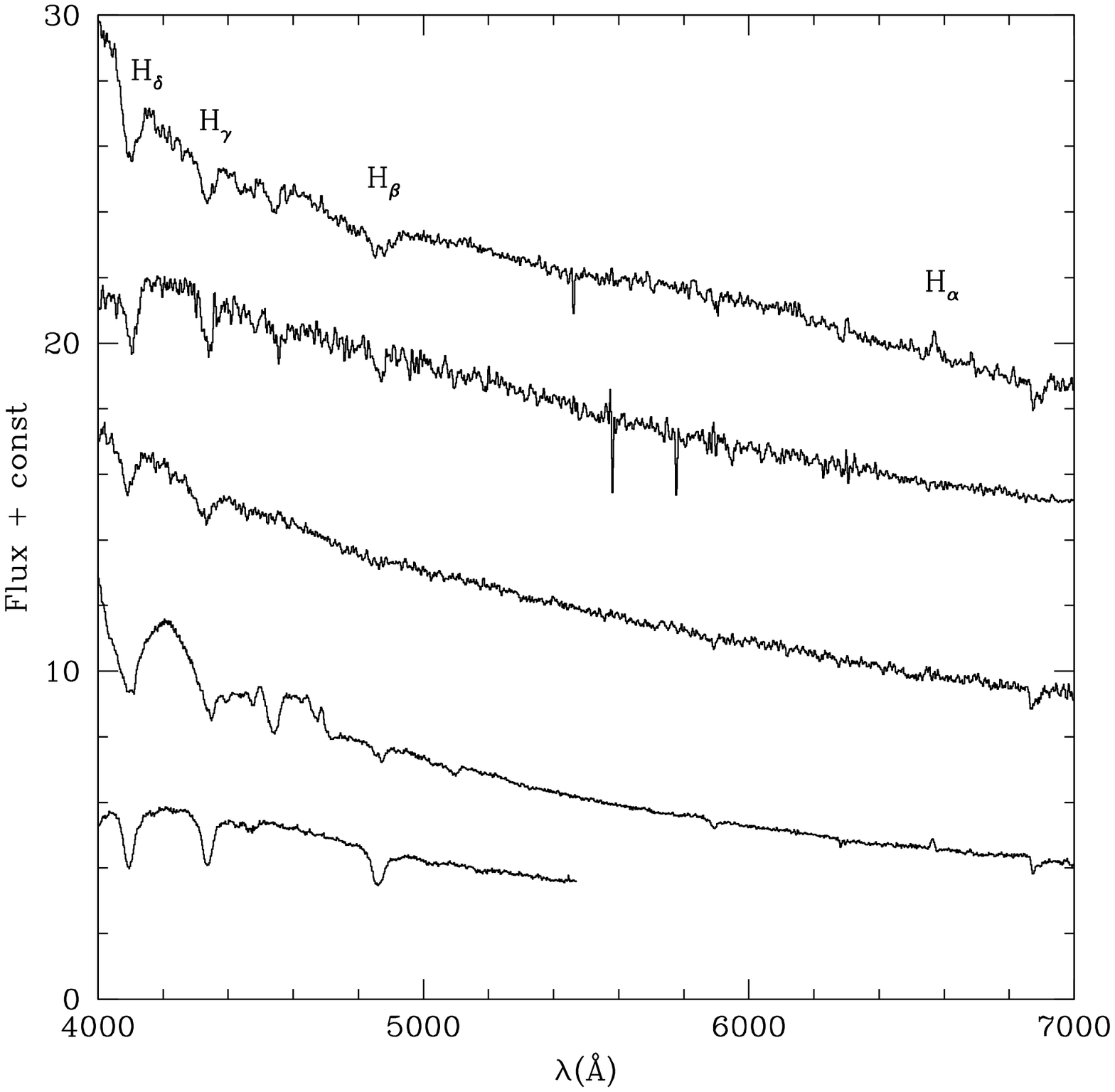}
\caption{\label{P200}
Palomar 5m spectra of CRTS cataclysmic variables.
In the left panel (top to bottom), we show spectra of three CVs (CSS100108:081031+002429, CSS110114:091937-055519, CSS100313:085607+123837)
exhibiting strong hydrogen and helium emission lines typical of CVs near quiescence. 
In the right panel (top to bottom), we show the spectra of CRTS CVs that were observed during outburst 
(CSS120113:040822+141516, CSS100507:164354-131525, CSS110623:173517+154708, CSS080606:162322+121334, CSS110501:094825+204333).
}
}
\end{figure*}

\begin{figure*}{
\includegraphics[width=84mm]{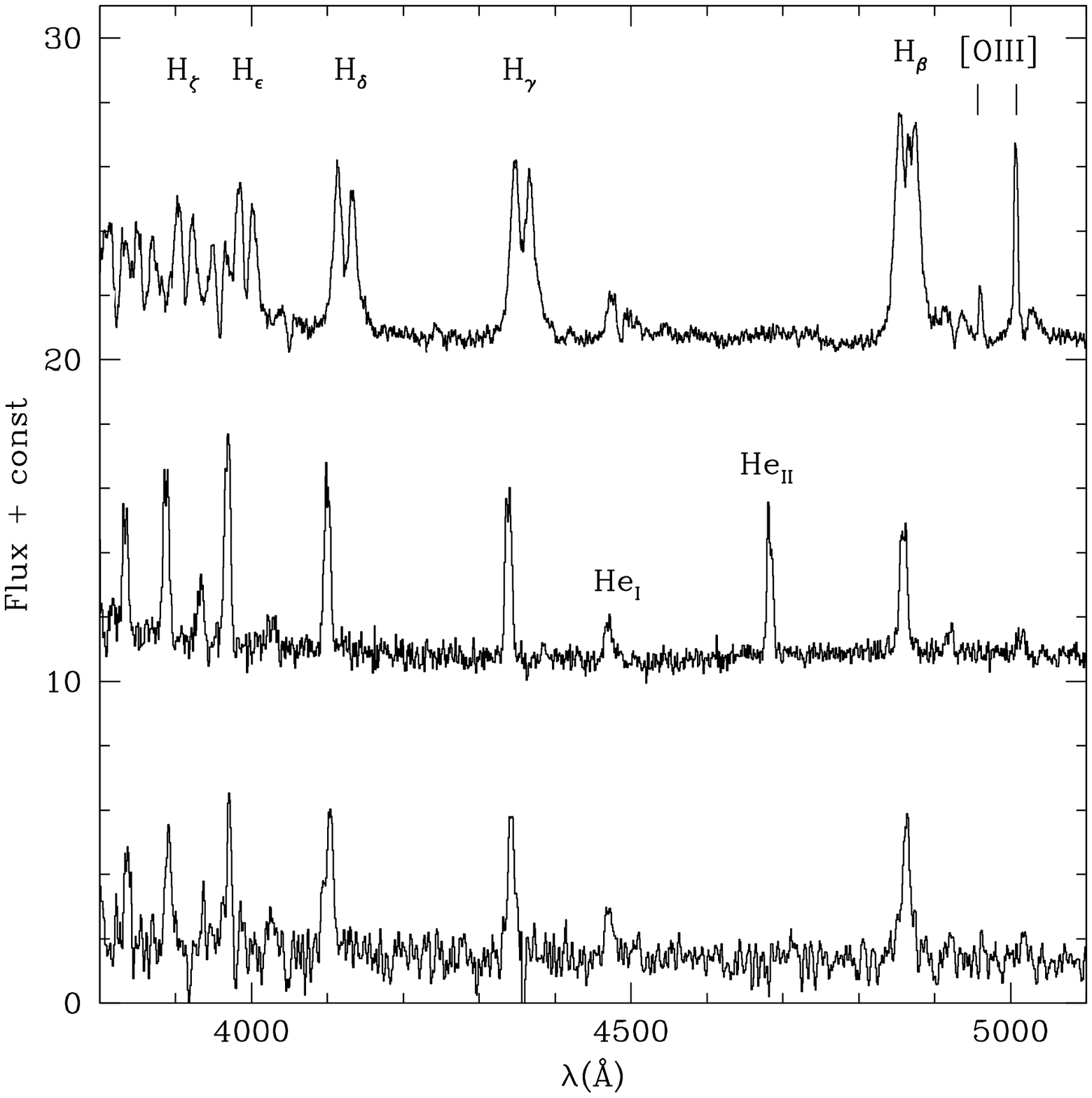}\includegraphics[width=84mm]{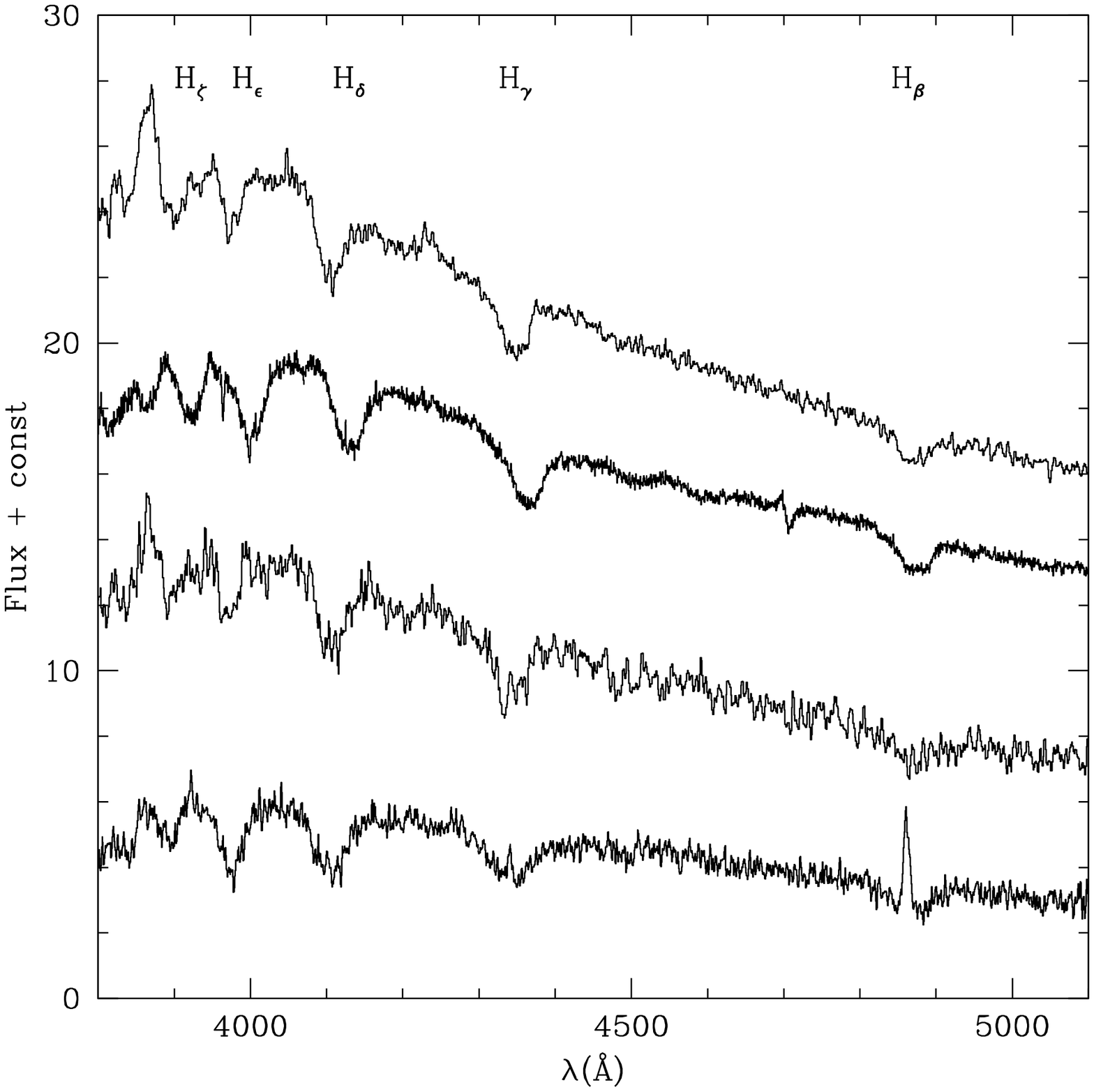}
\caption{\label{Keck}
Keck spectra of cataclysmic variables.
In the left panel (top to bottom), we show spectra for three CVs (CSS111003:054558+022106, CSS120526:165741-055625, CSS120613:212655-012054
with strong hydrogen and helium emission lines typical of CVs. CSS111003:054558+022106 lies within
the shell of a planetary nebula PN Te 11 (Jacoby et al.~2010), clearly seen within SDSS images,
and was found to have a period of 0.12 days by Miszalski et al.~(2011).
In the right panel (top to bottom), we show the spectra of four CRTS CVs that were observed during outburst (CSS120610:135419+273603, 
CSS120612:173245+094746, CSS111022:205145+075305, CSS100916:215226-012850).
}
}
\end{figure*}

We classified CRTS transients as CVs based upon the detection of strong, broad hydrogen absorption or emission lines
detected in our follow-up spectroscopy.  As many of the sources were observed shortly after they were detected in
outburst, the sample is biased toward systems exhibiting absorption lines from an optically thick accretion disk. This
is in contrast to spectroscopically confirmed CVs from SDSS, which are mainly systems observed during quiescence when
strong emission lines are observed.  In Figure \ref{P200}, we present examples of the Palomar DSBS spectra, and in
Figure \ref{Keck}, we present examples of Keck LRIS spectra of CRTS CVs.

\begin{figure}{
\includegraphics[width=84mm]{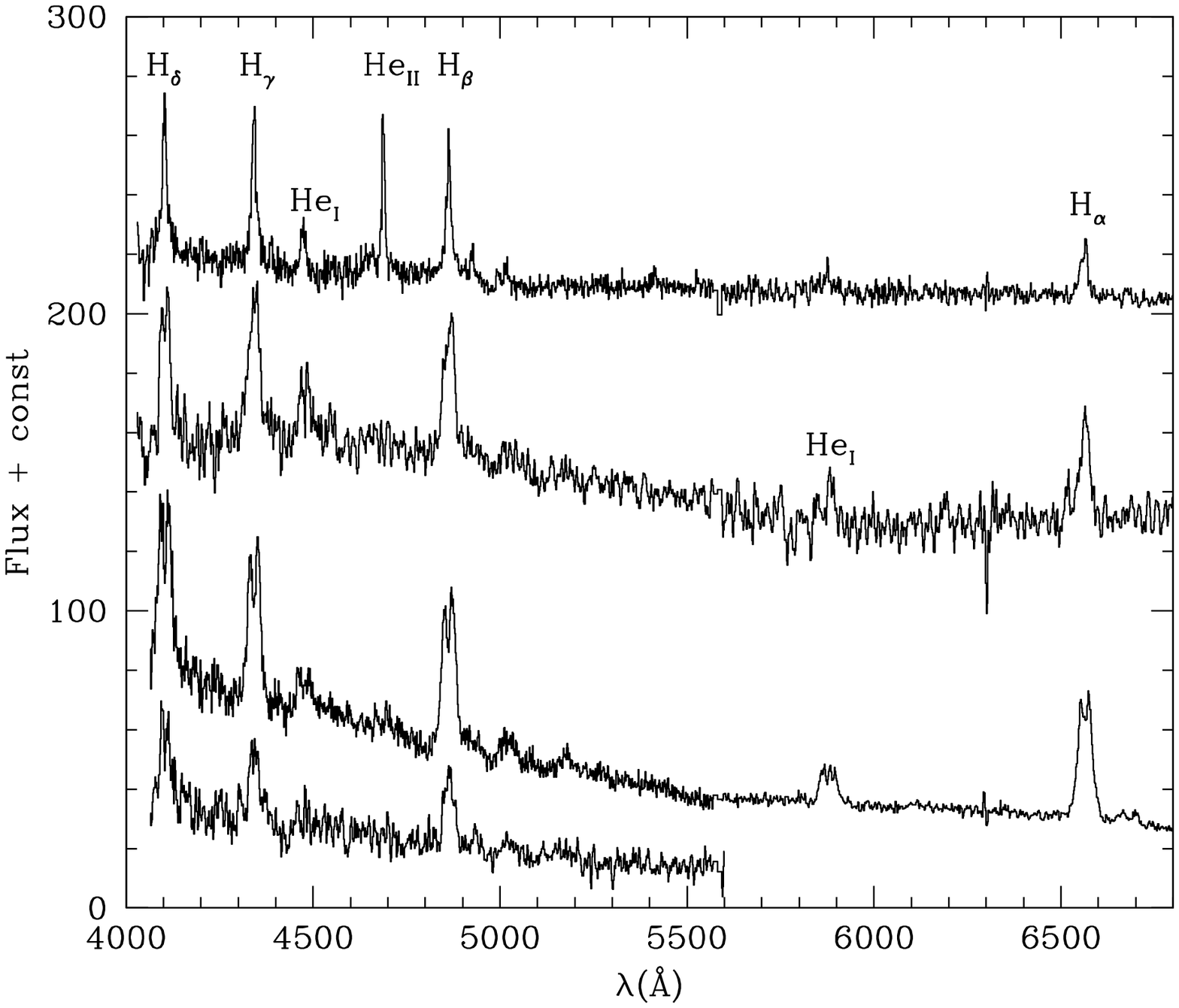}
\caption{\label{6dF}
The 6dF survey spectra of four cataclysmic variables.
From top to bottom, TY PsA (SSS100716:224940-270653), CC Scl (SSS111103:231532-304848),
YY Sex (CSS090301:103947-050658), and CSS120222:124602-202302.
}
}
\end{figure}

\begin{figure}{
\includegraphics[width=84mm]{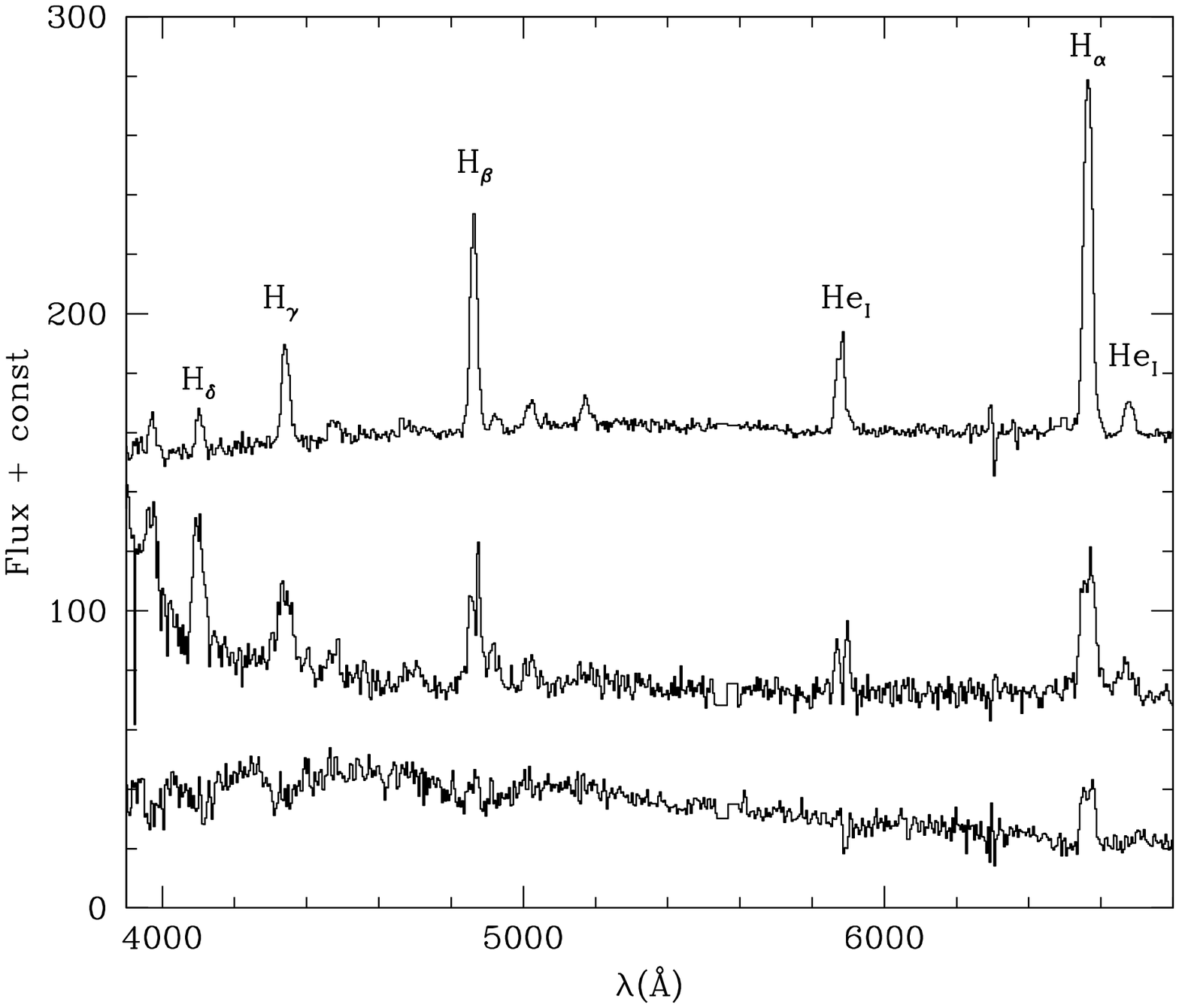}
\caption{\label{2dF}
The 2dF survey spectra of CRTS cataclysmic variables
From top to bottom, HV Vir (CSS080227:132103+015329), OU Vir (CSS080329:143500-004606) 
and SSS100911:222416-292422.
}
}
\end{figure}

We complemented our own follow-up observations with spectroscopy of four CRTS CV candidates observed by the Six-Degree Field
survey (6dF; Jones et al.~2004) DR3 spectra (Fig. \ref{6dF}). We also
matched the objects with the Two-Degree Field survey (2df; Croom et al. 2001), and found 
spectra for another three CRTS CVs (Fig. \ref{2dF}). The sources having SDSS spectra 
have already been analysed by Szkody et al.~(2002-2011).

\section{Magnetic CV systems}

Magnetic CVs form a subset of 10\% to 20\% of all known CVs (Wickramasinghe \& Ferrario 2000, Scaringi et al. 2010).
These objects are either polars (AM Her stars), or intermediate polars (IPs or DQ Her stars).  Intermediate polars have
weaker magnetic fields and may form partial accretion discs, while polars accrete directly from the secondary onto the
white dwarf.  In these systems, material is magnetically accreted onto the pole of the white dwarf giving rise to soft
(polars) or hard (IPs) X-rays that are modulated at the spin period of the white dwarf (Warner 2003). The detection of
X-ray flux is thus one means of separating IP and polar CV candidates from the more common dwarf novae.

\begin{table*}
\label{Rosat}
\caption{ROSAT matches to CRTS CVs}
\begin{minipage}{186mm}
\begin{center}
\begin{tabular}{@{}llllll}
\hline 
CRTS ID & HR1 & HR2 & Offset($\arcsec$) & N$\sigma$ &  ROSAT ID\\
\hline 
CRTS\_J001538.3+2636057 & 0.68 (0.44) & -0.25 (0.38) &  0.9 &  0.07 & 1RXSJ001538.2+263656$\ast$\\
CRTS\_J003203.6+3145010 & 1.00 (0.26) & 0.16 (0.41) & 29.8 &  1.30 & 1RXSJ003204.2+314441\\
CRTS\_J003304.0+3801006 & 0.55 (0.46) & 1.00 (0.48) & 23.9 &  0.99 & 1RXSJ003302.3+380118\\
CRTS\_J013308.7+3832017 & 1.00 (0.92) & 1.00 (9.99) &  2.5 &  0.18 & 1RXSJ013308.9+383218\\
CRTS\_J015051.5+3326022 & 1.00 (0.22) & 0.73 (0.18) &  3.4 &  0.29 & 1RXSJ015051.8+332622\\
CRTS\_J020804.2+3732017 & 1.00 (0.24) & 0.06 (0.39) & 27.8 &  1.39 & 1RXSJ020802.6+373236\\
CRTS\_J044027.1+0233001 & 1.00 (0.33) & 1.00 (0.30) &  2.2 &  0.20 & 1RXSJ044027.0+023300$\ast$\\
CRTS\_J051922.9+1554035 & 1.00 (0.31) & 0.37 (0.40) & 14.7 &  0.73 & 1RXSJ051922.6+155421\\
CRTS\_J053054.6-3357030 & -0.09 (0.36) & -0.66 (0.51) &  8.3 &  0.55 & 1RXSJ053054.5-335722\\
CRTS\_J065128.5-4109018 & 1.00 (0.39) & 0.20 (0.33) &  8.9 &  0.56 & 1RXSJ065127.9-410913\\
CRTS\_J073339.3+2122001 & 1.00 (0.58) & -0.51 (1.01) & 21.0 &  1.50 & 1RXSJ073340.7+212208\\
CRTS\_J094327.3-2720039 & 1.00 (0.40) & -0.14 (0.43) & 16.0 &  0.80 & 1RXSJ094326.1-272035$\ast$\\
CRTS\_J104411.4+2113007 & 0.69 (0.35) & 0.91 (0.72) & 37.1 &  1.95 & 1RXSJ104409.6+211334$\ast$ \\
CRTS\_J121924.7-1900024 & 0.58 (0.40) & 1.00 (2.40) & 12.1 &  1.01 & 1RXSJ121925.5-190022\\
CRTS\_J135143.5-4430020 & 0.33 (0.43) & 0.47 (0.57) & 14.6 &  0.91 & 1RXSJ135142.2-443023\\
CRTS\_J135915.2-3914052 & 0.57 (0.33) & -0.09 (0.44) &  6.4 &  0.53 & 1RXSJ135915.6-391447$\ast$\\
CRTS\_J152506.9-0326055 & 0.71 (0.20) & 0.38 (0.25) &  7.4 &  0.62 & 1RXSJ152506.9-032647$\ast$ \\
CRTS\_J172148.9-0517013 & 0.92 (0.27) & 0.73 (0.22) & 17.1 &  1.43 & 1RXSJ172148.4-051729\\
CRTS\_J203214.0-1126001 & 0.07 (0.38) & 0.33 (0.62) &  6.8 &  0.52 & 1RXSJ203214.0-112554\\
CRTS\_J221344.0+1732052 & 1.00 (0.21) & -0.32 (0.34) &  9.8 &  0.75 & 1RXSJ221343.9+173301$\ast$\\
CRTS\_J231909.2+3315040 & 0.95 (0.10) & 0.32 (0.23) &  9.9 &  0.76 & 1RXSJ231909.9+331544$\ast$\\
CRTS\_J233003.0+3033000 & 1.00 (0.49) & 1.00 (0.75) & 22.0 &  1.29 & 1RXSJ233004.7+303305$\ast$\\
\hline
\end{tabular}
\end{center}
\end{minipage}
\medskip 
Objects marked with `$\ast$' were identified as X-ray matches in CRTS alerts.
Col. (1) presents the CRTS identifier.
Cols. (2) and (3) present the ROSAT X-ray hardness ratios HR1 and HR2, respectively.
Col. (4) presents the offset between the CRTS and ROSAT sources.
Col. (5) presents the offset in terms of ROSAT position error $\sigma$.
Col. (6) presents the ROSAT identifier.
\end{table*}

To provide a more complete view of the X-ray properties of the CRTS CVs, we have matched the ROSAT X-ray catalog
positions (Voges et al.~1999, 2000). In total, 42 of the CVs match the position of ROSAT X-ray sources within $2\sigma$.
Of these, 22 are CRTS CVs and 20 are previously known CVs.  In Table 2, we present the ROSAT X-ray source matches to new
CRTS CVs.

During the course of the CRTS survey, most new optical transient discoveries were investigated using the Datascope
service\footnote{http://heasarc.gsfc.nasa.gov/vo}. This VO resource aggregates astronomical data covering a given
location, so that optical sources can be compared with prior detections in radio, X-ray, infrared and other wavelengths.
Sources that were found to match prior X-ray detections were noted on CRTS discovery webpages.  Our rapid alerting
system enabled optical follow-up of a number of the potential magnetic CV systems.  For example, Schwope \& Thinius
(2012) combined Catalina data with additional photometry and found that CSS091109:035759+102943 is a candidate polar
with an orbital period of 114 minutes.  Ten of the sources classified as CVs by CRTS were flagged as
possible X-ray sources (eg. Drake et al.~2009a). 

Among the X-ray sources followed, CSS090219:044027+023301 was found to be associated with 1RXS J044027+023300 and was
spectroscopically confirmed as a CV by Thorstensen \& Skinner (2012). CSS100217:104411+211307 was identified as a
possible match to 1RXS J104409.6+211334. However, this source was found to exhibit superhumps by Maehara et al.~(2010),
suggesting it is unlikely to be a magnetic system. In our CRTS Circular we noted CSS081231:071126+440405 to be an
apparent eclipsing polar-type
CV. This was found to be a magnetic low-accretion rate system by Thorne et al.~(2010). In contrast,
CSS120101:105123+672528 was found to match 1RXS J105120.5+672550 and was independently detected by the MASTER
project five hours after CRTS (Tiurina et al.~2012). Once again this was determined to be an SU UMa-type dwarf nova
(Pavlenko et al.~2012), rather than a magnetic system.  CSS120330:154450-115323, was also classified as a possible X-ray
match with ROSAT source 1RXS J15444.5-115340. As the nature of the object was unclear, we initially only classified 
the object as a variable star.  Optical spectroscopic follow-up observations along with ultraviolet and X-ray data revealed
the source to be a likely dwarf nova (Wagner et al.~2012), rather than a magnetic system.

\section{Eclipsing CV systems}

Cataclysmic variables that exhibit eclipses are relatively common. More than 150 such systems are listed in the
International Variable Star Index (VSX, Watson et al.~2006) database. These systems are particularly valuable as they
allow precise measurements of parameters such as orbital period (e.g. Littlefair et al. 2006, 2008; Savoury et al. 2011). Although
we have not undertaken a specific search for these objects in the photometric data, a number of clearly eclipsing CVs
were apparent from the lightcurves.

\begin{figure}{
\includegraphics[width=84mm]{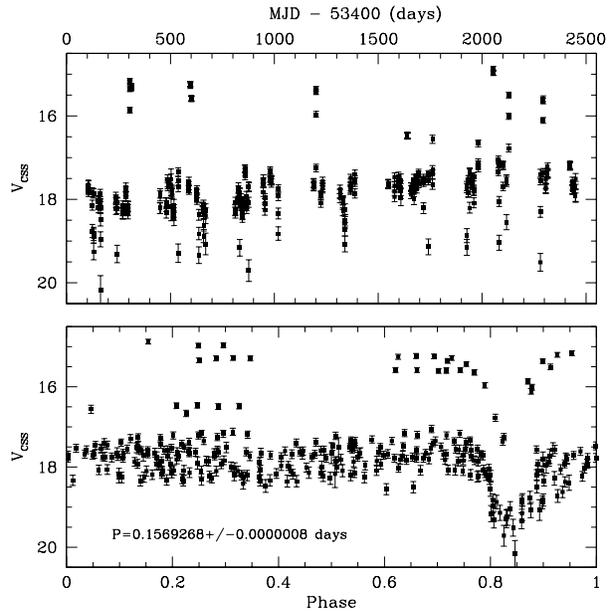}
\caption{\label{CSS110513}
The lightcurve of eclipsing CV CSS110513:210846-035031.
In the top panel we plot the observed lightcurve.
In the bottom panel we plot the lightcurve when folded
with the orbital period.
}
}
\end{figure}

CSS110513:210846-035031 was a CV that we classified as an eclipsing system.  This system shows outbursts up to $V \sim
15$ and eclipses to $V \sim 20$. In Figure \ref{CSS110513}, we plot the lightcurve of CSS110513:210846-035031 along with
the phase folded photometry. The eclipses of this object are observed over a span of $\sim 2000$ days. Although the
time of mid-eclipse is not
well defined, the long baseline allows determination of a very accurate period. For this system we find an eclipse
ephemeris of $MJD = 53500.131(5) + E \times 0.1569268(8)$. The figure clearly shows the eclipse both in the outburst and
quiescent photometry. The eclipse itself lasts $27 \pm 4$ mins.

Other eclipsing CVs include MLS120517:152507-032655, which we found to be a match to 
1RXS J152506.9-032647. This exhibits deep ($>2$ mag) eclipses, as do MLS101226:072033+172437, 
CSS090622:215636+193242, CSS080228:081210+040352, and CSS080227:112634-100210. The lightcurves
can be found on the CRTS website\footnote{http://crts.caltech.edu/}. Seven of the CVs detected 
by CRTS are previously known eclipsing systems.

\section{CV Periods}

We matched the CVs detected by CRTS with those in VSX and in Ritter \& Kolb (2003; July 2012 version), 
resulting in a sub-sample of 196 systems with known orbital period.

\begin{figure*}{
\includegraphics[width=84mm]{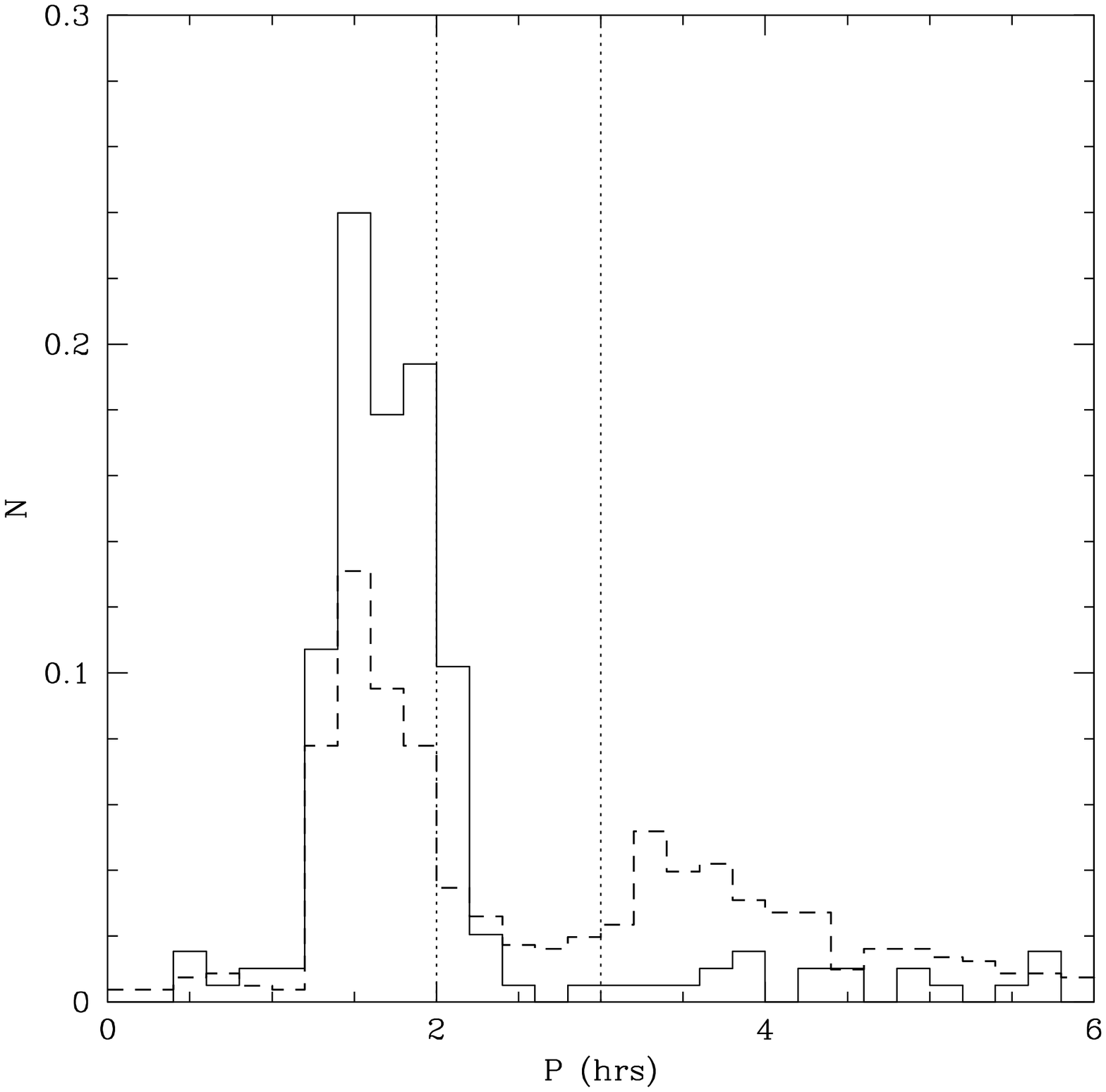}\includegraphics[width=84mm]{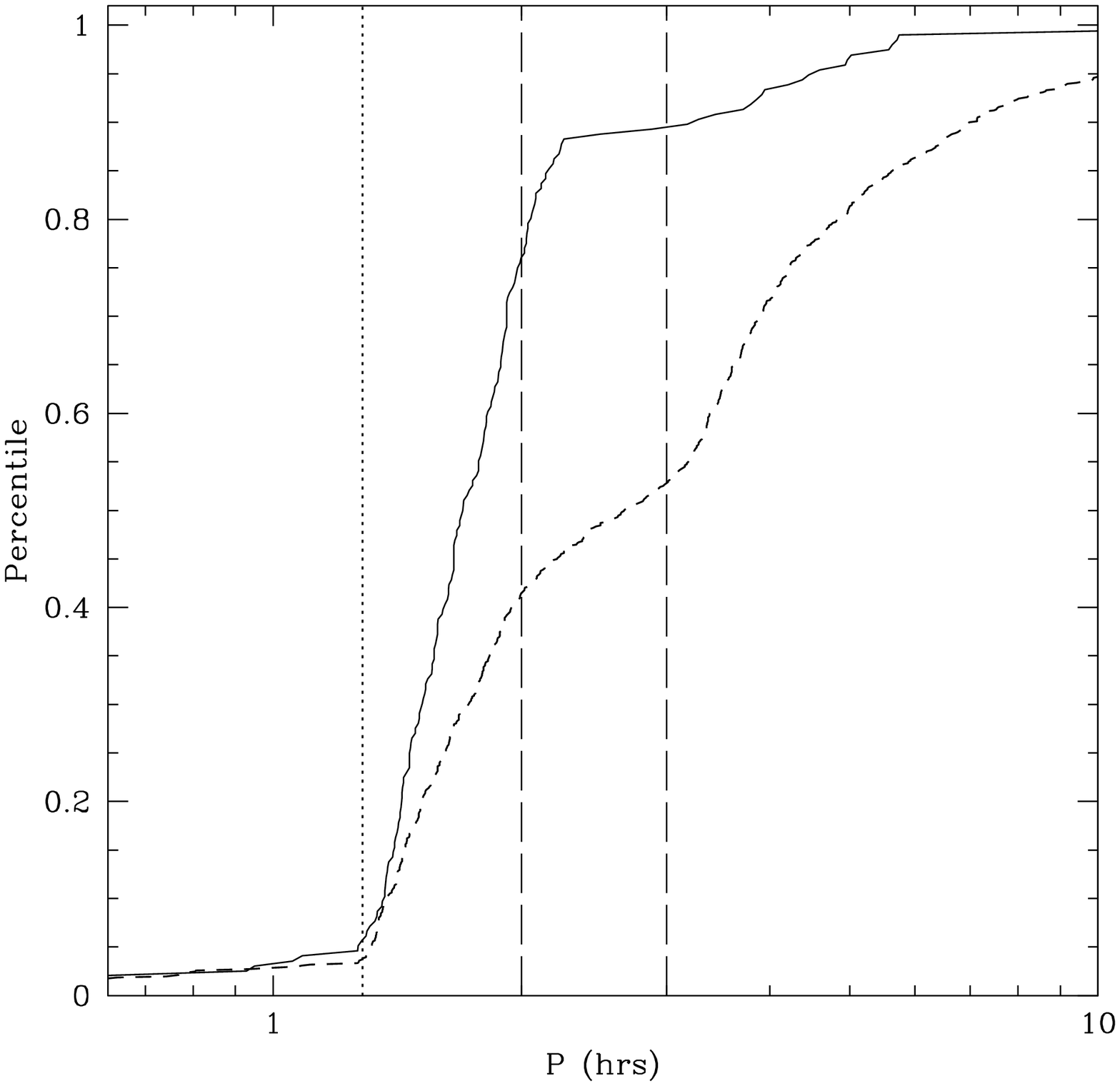}
\caption{\label{Per}
  The normalised CV period distribution.  On the left panel, we plot the distribution of CV periods for sources detected
  by CRTS (solid-line) and the distribution of all CV periods given by Ritter \& Kolb (2003, dashed-line).  On the right
  panel, we plot the cumulative distribution of CV periods. In the right panel CV, the period-minimum is plotted with a
  dotted line, while the period gap is indicated by the region between the two long-dashed lines.
}
}
\end{figure*}

In Figure \ref{Per} we plot the period distribution for the CVs detected by CRTS and the distribution for the 806 CVs in
the Ritter \& Kolb catalog that were not detected by CRTS.  For comparison we also plot the cumulative distributions.  A
similar analysis was performed by Woudt et al.~(2012). This work also demonstrated that the sources with shorter periods
had fewer outburst detections. However, the sample here is larger since Woudt et al.~(2012) did not include periods for
the bright CVs observed by VSNet that were also detected by CRTS.  This clearly shows that CRTS detects a much larger
fraction of CVs with short periods. A one-dimensional KS-test gives the probability that the two samples are drawn from
the same underlying distribution
of $P < 0.001$. 

\subsection{Ultra-short-period Systems}

In comparison to the full CRTS CV sample, a modest number have been found to be ultra-short-period systems that lie
below the orbital period minimum for normal CVs, which is near 80min (G\"ansicke et al. 2009). The dominant populations at such short orbital periods are AM CVn
systems, and SU UMa dwarf novae with substellar or evolved companions (e.g Carter et al. 2013; Levitan et al. 2011,
Thorstensen et al. 2002; Augusteijn et al. 1993; Littlefair et al. 2007; Uthas et al. 2011).  Woudt \& Warner (2010)
discovered that CSS111019:233313-155744 is an eclipsing dwarf nova with a 61.7-minute period while Woudt et al.~(2012)
found that CSS090331:102843-081927 is an ultra-short period CV (like V485 Cen and EI Psc), with a period of 52.1
minutes.  Breedt et al.~(2012) discovered that CSS100603:112253-111037 is a helium-rich dwarf nova with a 65-minute
orbital period. Interestingly, the authors noted that this object appears to be the first evidence for a dwarf nova
evolving into an AM CVn system.  Littlefield et al.~(2013) discovered that CSS120422:111127+571239 is a similar
hydrogen-depleted SU UMa-type CV below the period gap with a superhump period of 55.8 minutes.
The AMCVn system SDSS J172102.48+273301.2 (Rau et al.~2010) was also detected by CRTS in during outburst in July 2012.
Recently, Levitan et al.~(2013) spectroscopically confirmed that CSS090219:043518+002941 and CSS110507:163239+351108
are AM CVn systems, yet did not determine their orbital periods.

\section{CV Distances}

The absolute magnitudes of CVs are well known to vary significantly with their orbital periods. Short-period CVs
typically have $M_V \sim 9.5$ in quiescence and $M_V \sim 5$ in outburst (Warner et al.~1987).  However, recent
determinations of distances suggest that quiescent magnitudes near the period minimum can be much fainter, at $M_g=11.6$
(G\"ansicke et al.~2009), or perhaps even as faint as $M_V=14$ (Sproats et al.~1996), and it is those type 
of systems which are likely to dominated the galactic CV population.
In contrast, the determinations of CV outburst absolute magnitudes have remained relatively consistent with the values
predicted by Warner (1987) (eg.  Thorstensen 2003, Harrison et al. 2004).  The outburst absolute magnitudes are much
better constrained than those in quiescence since the accretion disc completely dominates the emission during outburst.
The brightness scales with the physical size of the disc, which itself scales with the orbital period.  As the majority
of CRTS CVs have periods around 1.7 hours, if one was to assume the objects have outburst magnitudes that are consistent 
with the prescription of Warner (1987), the absolute magnitudes of most are expected to be $M_V(\rm max) \sim 5.5$.

\begin{figure*}{
\includegraphics[width=84mm]{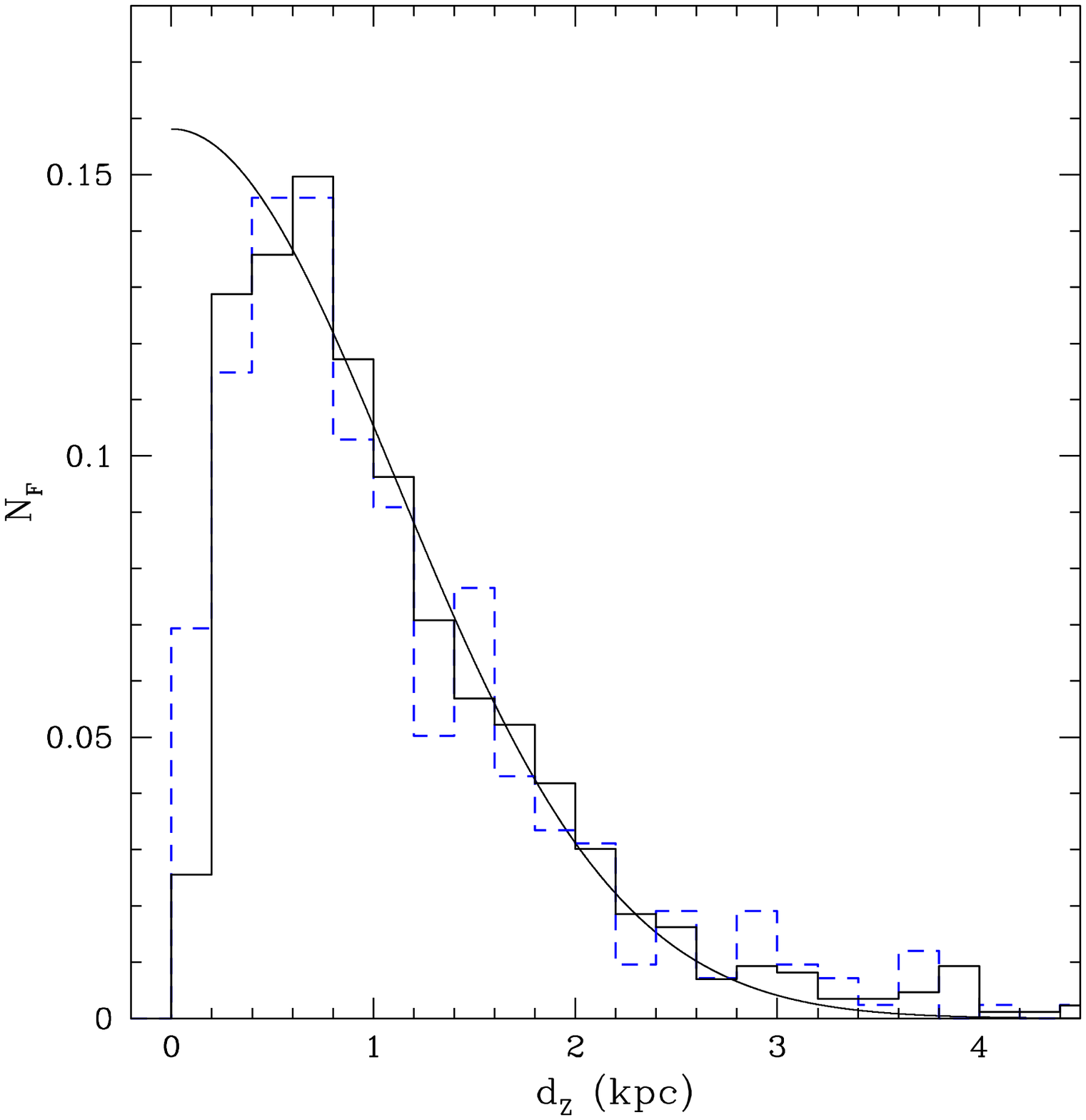}\includegraphics[width=84mm]{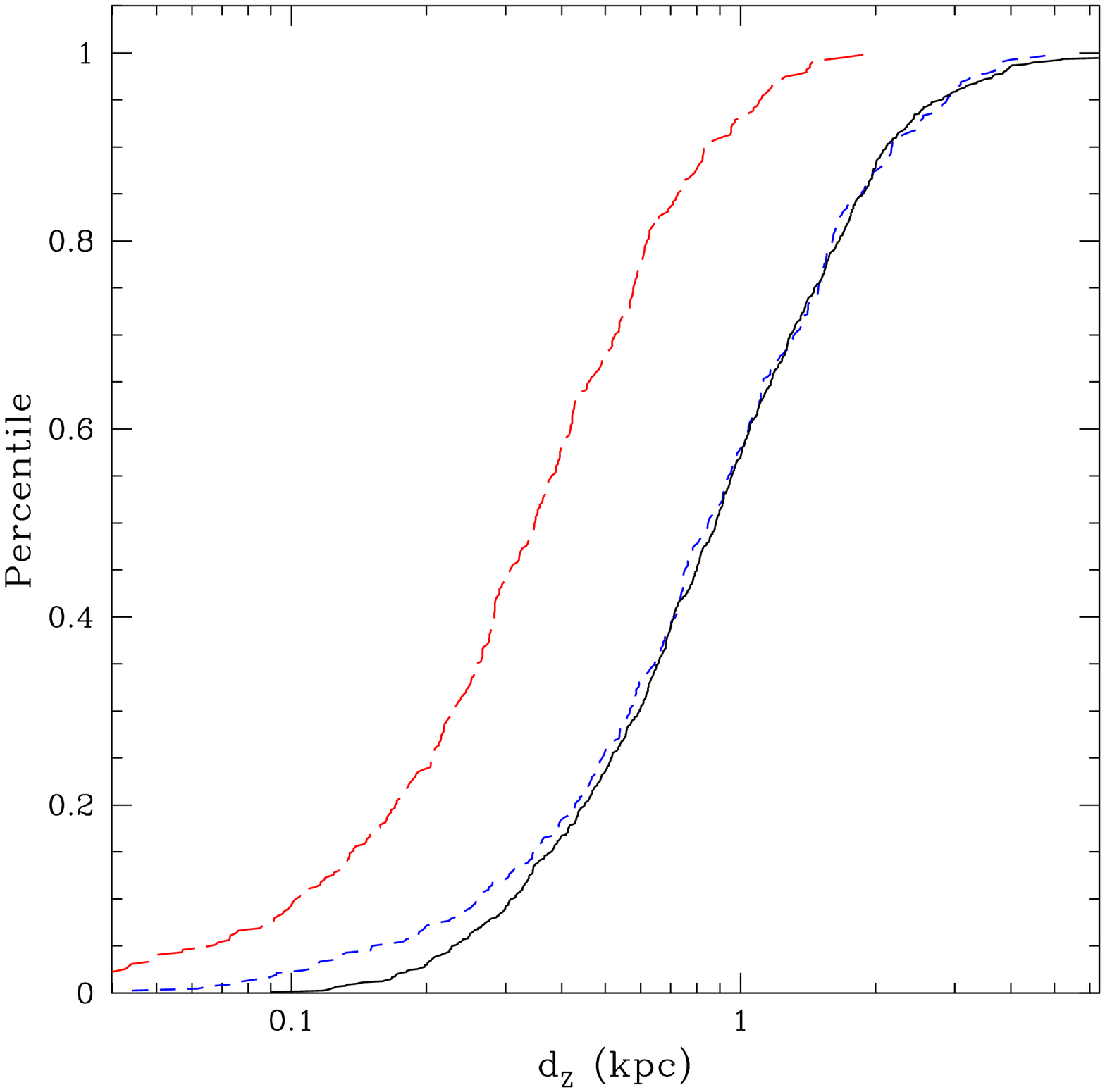}
\caption{\label{dist}
    Normalised distribution of CV heights above the Galactic mid-plane.  In the left panel we show the distances derived from
  391 SDSS magnitudes assuming quiescent magnitudes of $Mg(\rm min) = 9.5$ as a dashed blue line. For the 855 CRTS CVs we plot
  the distances assuming $M_V(\rm max) = 5.5$ (solid black line). Gaussian fits to the distributions give a scale height
  of 1.1kpc.  In the right panel we plot the cumulative distribution of distances derived from SDSS magnitudes assuming
  $Mg(\rm min) = 9.5$ (blue short-dashed line), the CRTS peak outburst magnitudes assuming $M_V(\rm max) = 5.5$ (black solid-line),
  and from the SDSS magnitudes adopting $Mg(\rm min) = 11.6$ (red long-dashed line).
}
}
\end{figure*}

In Figure \ref{dist}, we plot the distribution of CV distances. The agreement between distances derived from CRTS
outburst magnitudes and SDSS quiescent magnitudes (assuming $M_g(\rm min)=9.5$) is very good. For the two distributions we
find scale heights of 1.1 kpc. This value is consistent with values for the scale height of the thick disc (eg. Carollo
et al.~2010; Du et al. 2006; Larsen \& Humphreys 2003; Kerber et al. 2001) and suggests that the CVs are drawn from this
distribution.  However, if one adopts $M_g(\rm min)=11.6$ as found by G\"ansicke et al.~(2009), the scale height is reduced
to 420pc. This value is still significantly larger than most values derived for the Galactic thin disc (eg. Gould et
al. 1997; Chen et al. 2001), but consistent with some results (eg. Nelson et al.~2002).

Considering the survey is largely limited to sources with $|b| > 10\arcdeg $, the faintest CRTS CVs are preferentially
found at greater scale heights.  Nevertheless, even with this bias there should be few thin disc sources beyond three
times the scale height. However, we note that the distances derived using quiescent absolute magnitudes are clearly
inconsistent with the distances derived from outburst magnitudes. Both sets of distances are inconsistent with CVs
coming from a population with a scale height of 190pc, as found by Patterson~(1984). This result is not surprising since
the Patterson~(1984) sample is based on nearby CVs. Only recent all-sky surveys are deep and wide enough to probe CV
systems at thick disc distances.  More recently, Pretorius et al.~(2007) considered a number of models of the CV
distribution within the Galaxy and suggested that, given the different ages of the components of the Galaxy, the scale
heights of CVs should depend on their age, and hence on their orbital period.  From their model {\it A1} they found that
CVs with $|b| > 20 \arcdeg $ would be strongly concentrated toward the period minimum when sources with $V > 20$ were
considered. The CRTS period distribution appears consistent with this model.  Pretorius et al.~(2007) also suggest that
the flux-limited samples available at that time were not deep enough to reproduce the intrinsic population CVs, and thus
that any comparison between the observed population and true population had to remain difficult. Nevertheless, the
determination of the spatial density of CVs is an important means for understanding CV evolution. In a future
paper we shall address this question for CVs detected by CRTS (Breedt et al.~2014).

\section{Discussion}

In this work we have cataloged 855 CV candidates detected by CRTS. Of these 705 are new discoveries and at least 137
are spectroscopically confirmed.  These sources were primarily selected among CRTS optical transient sources 
using information about prior outbursts and their amplitudes, combined with archival information such as optical
colours, and the presence of radio and X-ray sources. We have investigated the resulting outburst amplitude and 
colour distributions and confirm the expected differences from other common types of optical transients (such as 
supernovae and blazars). We find that the CRTS CV sample extends two magnitudes deeper compared to the CVs that 
were spectroscopically identified by SDSS (Szkody 2002-2011).

In contrast to the recent work on quantifying the CV population based on CRTS data by Thorstensen \& Skinner (2012), our
analysis includes the additional CVs discovered using MLS and SSS telescopes, as well as more accurate details of the
CRTS CV detection and classification. Nevertheless, given that only 45\% of the CRTS CVs have been detected in outburst
more than once, we agree with the suggestion of Thorstensen \& Skinner (2012) that a large fraction of the galactic CV
population must remain to be discovered.

While it is clear that the Galactic latitude limits of the Catalina survey ($|b| > 10\arcdeg $) create a bias towards
sources at large scale heights, the absolute magnitudes predicted from CV outbursts suggest a significant fraction of
the CVs discovered have scale heights well beyond that expected for the Galactic thin disc. This strongly supports the
idea that many of the CRTS CVs belong to a thick disc population.

Since CRTS detects CVs by searching for optical transient sources, this inherently biases the detections toward dwarf
nova systems.  To better understand the CV population we identified candidates that coincide with hard sources from X-ray
catalogs. We find that most of the CV systems with X-ray matches appear to be non-magnetic systems. Nevertheless, CRTS
has discovered both eclipsing systems and magnetic CVs. It is likely that more magnetic CVs remain to be identified.

By analysing the orbital periods of the CVs, we find that the period distribution of CRTS CVs includes a much more
significant contribution from short-period systems compared to the bulk of the previously known CVs, which are, on
average, both brighter, and have more frequent outbursts. This result is in agreement with prior analyses by Wils et
al.~(2010) and Woudt et al.~(2012), and also with the work by G\"ansicke et al.~(2009) based on SDSS CVs, and Uemura et
al.~(2010) on WZ Sge stars.  This underlines that new CV samples that have much deeper limiting magnitudes probe a
different population of systems compared to the previously discovered bright CVs.  However, in contrast to the orbital 
period distribution, we find no evidence for a difference in the outburst rates of the dwarf nova CVs.

Recent work on CVs from CRTS and other synoptic surveys has led to the discovery of a number of ultra-short-period CVs.
These have been identified as AM CVn types (eg. Woudt \& Warner 2010; Levitan et al.~2013) as well as those that systems
are evolving into AM CVns (eg. Breedt et al.~2012), or have substellar companions (eg. Garnavich et al.~2012).  Given
that less than quarter of the CRTS CVs currently have orbital period determinations, it is likely that a large number of
ultra-short-period systems remain to be found.

\section*{Acknowledgments}

CRTS and CSDR1 are supported by the U.S.~National Science Foundation under grants AST-0909182 and CNS-0540369.  The work
at Caltech was supported in part by the NASA Fermi grant 08-FERMI08-0025, and by the Ajax Foundation. The CSS survey is
funded by the National Aeronautics and Space Administration under Grant No. NNG05GF22G issued through the Science
Mission Directorate Near-Earth Objects Observations Program. J. L. P. acknowledges support from NASA through Hubble
Fellowship Grant HF-51261.01-A awarded by the STScI, which is operated by AURA, Inc.  for NASA, under contract NAS
5-26555.  Support for M.C. and G.T. is provided by the Ministry for the Economy, Development, and Tourism's Programa
Iniciativa Cient\'{i}fica Milenio through grant P07-021-F, awarded to The Milky Way Millennium Nucleus; by Proyecto
Basal PFB-06/2007; and by Proyecto FONDECYT Regular \#1110326.  SDSS-III is managed by the Astrophysical Research
Consortium for the Participating Institutions of the SDSS-III Collaboration Funding for SDSS-III has been provided by
the Alfred P. Sloan Foundation, the Participating Institutions, the National Science Foundation, and the U.S. Department
of Energy Office of Science. The SDSS-III web site is http://www.sdss3.org/.  This research has made use of the
International Variable Star Index (VSX) database, operated at AAVSO, Cambridge, Massachusetts, USA The Virtual
Astronomical Observatory (VAO) is managed by the VAO, LLC, a non-profit company established as a partnership of the
Associated Universities, Inc. and the Association of Universities for Research in Astronomy, Inc. The VAO is sponsored
by the National Science Foundation and the National Aeronautics and Space Administration.  The research leading to these
results has received funding from the European Research Council under the European Union's Seventh Framework Programme
(FP/2007-2013) / ERC Grant Agreement n. 320964 (WDTracer).  BTG was supported in part by the UK's Science and Technology
Facilities Council (ST/I001719/1).

\newpage

\end{document}